\documentstyle[12pt]{article}
 
\def\Qslash{Q\kern-0.15em\raise0.17ex\llap{/}\kern0.15em\relax} 
\def\hatQslash{\hat{Q}\kern-0.15em\raise0.17ex\llap{/}
\kern0.15em\relax} 
\def\Kslash{K\kern-0.15em\raise0.17ex\llap{/}\kern0.15em\relax} 
\def\Pslash{P\kern-0.15em\raise0.17ex\llap{/}\kern0.15em\relax} 
\def\Nslash{N\kern-0.15em\raise0.17ex\llap{/}\kern0.15em\relax} 
\def\Rslash{R\kern-0.1em\raise0.3ex\llap{/}\kern0.15em\relax} 
\def\appendix{%
\section*{Appendix}
 \setcounter{equation}{0}
 \setcounter{section}{1}
 \def\theequation{\mbox{\Alph{section}.\arabic{equation}}}
 }
\begin{document}
\hspace*{1ex}

\mbox{\hspace*{58ex}} OCU-PHYS-162 

\mbox{\hspace*{59ex}} October 1 1996 

\vspace*{10mm}

\begin{center}
{\Large {\bf 
Self-consistent determination of hard modes in hot QCD}} 
\end{center}

\hspace*{3ex}

\hspace*{3ex}

\hspace*{3ex}

\begin{center} 
{\large {\sc A. Ni\'{e}gawa}\footnote{ 
E-mail: niegawa@sci.osaka-cu.ac.jp}

{\normalsize\em Department of Physics, Osaka City University } \\ 
{\normalsize\em Sumiyoshi-ku, Osaka 558, Japan} } \\
\end{center} 

\hspace*{2ex}

\hspace*{2ex}

\hspace*{2ex}

\hspace*{2ex}
\begin{center} 
{\large {\bf Abstract}} \\ 
\end{center} 
\begin{quotation}
We determine self-consistently the hard-quark and hard-gluon 
modes in hot QCD. 
The damping-rate part in resummed hard-quark or hard-gluon 
propagators, rather than the thermal-mass part, plays the dominant 
role. 
\end{quotation}
\newpage
\setcounter{equation}{0}
\setcounter{section}{0}
\section{Introduction} 
\def\theequation{\mbox{\arabic{section}.\arabic{equation}}}
It has been established by Pisarski and Braaten 
Since it has been realized that, within the hard-thermal-loop (HTL) 
resummation scheme \cite{pis1,bra1} of perturbative hot QCD, the 
damping rate for a moving {\em particle} diverges logarithmically, 
the damping rate has continuously attracted much interest 
\cite{pis1,damp,smilga,pis5}. The {\em particle} (quark or gluon) is 
a \lq\lq good'' or stable mode in vacuum QCD. Then, the 
above-mentioned diverging damping rate indicates that, at nonzero 
temperature ($T \neq 0$), this particle \lq\lq damps'' 
instantaneously. 

Landsman has pointed out \cite{landsman} that the particles in 
vacuum theory are {\em not} \lq\lq good'' modes in thermal field 
theories. On the basis of a group-theoretical analysis, he then has 
proposed a notion of a non-shell particles as \lq\lq good'' or in a 
sense stable modes at $T \neq 0$. On the other hand, Umezawa and his 
coworkers have introduced \cite{umezawa} \lq\lq thermal 
quasiparticles''. In both approaches, the \lq\lq good'' modes are 
designed to be determined essentially in self-consistent manners. As 
to the soft modes, the HTL-resummed effective propagators 
\cite{pis1,bra1} summarize the \lq\lq good'' modes. For hard modes, 
although not fully comprehensive, studies along this line have been 
pursued, e.g., in \cite{smilga,umezawa,henn}. 

The purpose of this paper is to determine the \lq\lq good'' hard 
modes ($Q^\mu$ $=$ $O (T)$) to leading order at logarithmic accuracy 
within HTL-resummation scheme of perturbative hot QCD. By \lq\lq 
logarithmic accuracy'' we mean that the factor of $O \{ 1 / \ln 
(g^{- 1}) \}$ is ignored when compared to the factor of $O (1)$. We 
work in massless SU(N) \lq\lq QCD'' with $N_f$ quarks. 
\setcounter{equation}{0}
\setcounter{section}{1}
\section{Preliminary} 
We start with defining the quasifree Lagrangian density for the 
\lq\lq good'' modes, 
\begin{eqnarray} 
{\cal L}_0 & = & {\cal L}_0^{(q)} + {\cal L}_0^{(g)} + 
{\cal L}_0^{(FP)} \, , 
\label{2.1a} \\ 
{\cal L}_0^{(q)} & = & \bar{\psi} \left[ i 
{\partial\kern-0.0em\raise0.3ex\llap{/}\kern0.15em\relax} - \Sigma_F 
(i \partial) \right] \psi \, , 
\label{qf-quark} \\ 
{\cal L}_0^{(g)} & = & - \frac{1}{4} \left( \partial_\mu A^a_\nu - 
\partial_\nu A^a_\mu \right) \left( \partial^\mu A^{a \nu} - 
\partial^\nu A^{a \mu} \right) \nonumber \\ 
& & + \frac{1}{2} \, A^a_\mu \, \Pi_F^{\mu \nu} (i \partial) \, 
A^a_\nu - \frac{1}{2 \eta} \, \left( \partial^\mu A^a_\mu \right) 
\left( \partial^\nu A^a_\nu \right) \, , 
\label{qf-gluon} \\ 
{\cal L}_0^{(\mbox{\scriptsize{FP}})} & = & \left( \partial^\mu 
\bar{\eta}^a \right) \left( \partial_\mu \eta^a - g f^{a b c} 
A_\mu^b \eta^c \right) - \bar{\eta}^a \Pi^g_F (i \partial) \eta^a 
\, , 
\label{qf-FP} 
\end{eqnarray} 
where \lq\lq FP'' stands for Faddeev-Popov ghost field. $\Sigma_F 
(i \partial)$ in (\ref{qf-quark}) is a $4 \times 4$ matrix function 
of $i \partial$, which may be decomposed as 
\begin{equation} 
\Sigma_F (Q) = f (Q) \, 
{Q\kern-0.1em\raise0.3ex\llap{/}\kern0.15em\relax} + g (Q) \, 
\gamma^0 \, . 
\label{fg-def} 
\end{equation} 
Similarly $\Pi_F^{\mu \nu} (i \partial)$ in (\ref{qf-gluon}) may be 
decomposed as \cite{lan} 
\begin{eqnarray} 
\Pi_F^{\mu \nu} (Q) & = & {\cal P}_T^{\mu \nu} (Q) \, \Pi_F^T (Q) + 
{\cal P}_L^{\mu \nu} (Q) \, \Pi_F^L (Q) \nonumber \\ 
& & + \, {\cal C}^{\mu \nu} (Q) \, \Pi_F^C (Q) + {\cal D}^{\mu \nu} 
(Q) \, \Pi_F^D (Q) \, . 
\label{2-1} 
\end{eqnarray} 
Here 
\begin{eqnarray} 
{\cal P}_T^{\mu \nu} (Q) & \equiv & - \sum_{i, \, j = 1}^3 g_{\mu i} 
\, g_{\nu j} [\delta^{i j} - \hat{q}^i \hat{q}^j] 
\label{pt} \\ 
{\cal P}_L^{\mu \nu} (Q) & \equiv & g^{\mu \nu} - \frac{Q^\mu 
Q^\nu}{Q^2} - {\cal P}_T^{\mu \nu} (Q) \, , 
\label{pl} \\ 
{\cal C}^{\mu \nu} (Q) & \equiv & \frac{1}{\sqrt{2} q_0 q} \left[ 
Q^\mu \tilde{Q}^\nu + Q^\nu \tilde{Q}^\mu + 2 q^2 
\frac{Q^\mu Q^\nu}{Q^2 + i 0^+}  \right] \, , 
\label{c} \\  
{\cal D}^{\mu \nu} (Q) & \equiv & \frac{Q^\mu Q^\nu}{Q^2 + i 0^+} 
\, , 
\label{d} 
\end{eqnarray} 
where $\hat{{\bf q}} \equiv {\bf q} / q$ with $q \equiv |{\bf q}|$ 
and $\tilde{Q}^\mu \equiv (0, {\bf q})$. ${\cal P}_T^{\mu \nu} (Q)$ 
[${\cal P}_L^{\mu \nu} (Q)$] is the projection operator onto the 
transverse [longitudinal] mode. As is well known \cite{lan}, BRS 
invariance of the full QCD Lagrangian leads to (see below) 
\begin{eqnarray} 
\Pi^D_F (Q) = 0. 
\label{piD0} 
\end{eqnarray} 

Here it is worth making the following remark. As in \cite{umezawa}, 
${\cal L}_0$ in (\ref{2.1a}) - (\ref{qf-FP}) is non-hermitian, since 
$\Sigma_F$, $\Pi_F^{\mu \nu}$, and $\Pi_F^g$ are complex functions. 
Through a standard procedure, the quasifree Hamiltonian, $H_0$, is 
constructed from (\ref{2.1a}) - (\ref{qf-FP}), which is also 
non-hermitian. We recall that, in constructing the Gell-Mann-Low 
formula of perturbation theory in vacuum theory, the hermiticity of 
the free Hamiltonian plays an essential role. In the operator 
formalism of thermal field theory, which is called thermo field 
dynamics \cite{umezawa}, the so-called hat-Hamiltonian, $\hat{H}$, 
plays the role of Hamiltonian, $H$, in vacuum theory. $\hat{H}$ is 
defined as $\hat{H}$ $=$ $H$ $-$ $\tilde{H}$, where $\tilde{H}$ is 
constructed from $H$ through the so-called tilde-conjugation rules. 
The Gell-Mann-Low formula may be derived \cite{umezawa} by choosing 
a free Hamiltonian, $H_0$, from which the hat-Hamiltonian, 
$\hat{H}_0$ $=$ $H_0$ $-$ $\tilde{H}_0$, is constructed. It should 
be stressed that $H_0$ is not necessarily hermitian. In the course 
of derivation, the so-called tildicity of $\hat{H}_0$, i.e., the 
invariance of $- i \hat{H}_0$ under the tilde conjugation, plays the 
role of hermiticity of $H_0$ in vacuum theory. It is well known 
that, as far as thermal-equilibrium cases are concerned, both the 
above operator formalism and the conventional real-time thermal 
field theory (constructed on a time-path in a complex time plane) 
lead to the same Feynman rules in perturbative calculation. 

The interaction Lagrangian density is defined as 
\begin{equation} 
{\cal L}_{\mbox{\scriptsize int}} = 
{\cal L}_{\mbox{\scriptsize QCD}} - {\cal L}_0 \, . 
\label{int-1} 
\end{equation} 

On the basis of the theory defined by (\ref{2.1a}) - (\ref{qf-FP}) 
and (\ref{int-1}), we shall determine $\Sigma_F (Q)$, 
$\Pi_F^{\mu \nu} 
(Q)$, and $\Pi_F^g (Q)$ self consistently, to leading order at 
logarithmic accuracy. We employ the closed-time-path formalism of 
real-time thermal field theory \cite{lan}. 

The diagonalized or Feynman propagator of the quark constructed from 
(\ref{qf-quark}) and (\ref{fg-def}) is 
\[ 
\displaystyle{ \raisebox{1.1ex}{\scriptsize{$\diamond$}}} 
\mbox{\hspace{-0.33ex}} S_F (Q) = - \frac{1}{2} \sum_{\tau = \pm} 
\hat{{Q\kern-0.1em\raise0.3ex\llap{/}\kern0.15em\relax}}_\tau \, 
\frac{1}{\displaystyle{ \raisebox{1.1ex}{\scriptsize{$\diamond$}}} 
\mbox{\hspace{-0.33ex}} D_\tau (Q)} \, , 
\] 
where 
\begin{eqnarray} 
\hat{Q}^\mu_\tau & \equiv & (1, \tau \hat{{\bf q}}) \, , \nonumber 
\\ 
\displaystyle{ \raisebox{1.1ex}{\scriptsize{$\diamond$}}} 
\mbox{\hspace{-0.33ex}} D_\tau (Q) & = & (- q_0 + \tau q) \{ 1 - 
f (Q) \} + g (Q) \, . 
\label{dia-D} 
\end{eqnarray} 
Each component of the $2 \times 2$ matrix propagator is obtained 
from $\displaystyle{ \raisebox{1.1ex}{\scriptsize{$\diamond$}}} 
\mbox{\hspace{-0.33ex}} S_F (Q)$ through Bogoliubov transformation 
\cite{lan}: 
\begin{equation} 
\displaystyle{ \raisebox{1.1ex}{\scriptsize{$\diamond$}}} 
\mbox{\hspace{-0.33ex}} S^{(j i)} (Q) = \sum_{\tau = \pm} 
\hat{{Q\kern-0.1em\raise0.3ex\llap{/}\kern0.15em\relax}}_\tau 
\displaystyle{ \raisebox{1.1ex}{\scriptsize{$\diamond$}}} 
\mbox{\hspace{-0.33ex}} \tilde{S}^{(j i)}_\tau (Q) \, , 
\;\;\;\;\;\;\;\; (j, \, i = 1, 2) \, , 
\label{q-pro-1}  
\end{equation} 
where $i$ and $j$ are the thermal indexes and 
\begin{eqnarray} 
Re \, \displaystyle{ \raisebox{1.1ex}{\scriptsize{$\diamond$}}} 
\mbox{\hspace{-0.33ex}} \tilde{S}^{(1 1)}_\tau (Q) & = & - Re \, 
\displaystyle{ \raisebox{1.1ex}{\scriptsize{$\diamond$}}} 
\mbox{\hspace{-0.33ex}} \tilde{S}^{(2 2)}_\tau (Q) \nonumber \\ 
& = & - \frac{1}{2} \, Re \, \frac{1}{\displaystyle{ 
\raisebox{1.1ex}{\scriptsize{$\diamond$}}} \mbox{\hspace{-0.33ex}} 
D_\tau (Q)} \, , 
\label{ReS11} \\
Im \, \displaystyle{ \raisebox{1.1ex}{\scriptsize{$\diamond$}}} 
\mbox{\hspace{-0.33ex}} \tilde{S}^{(1 1)}_\tau (Q) & = & Im \, 
\displaystyle{ \raisebox{1.1ex}{\scriptsize{$\diamond$}}} 
\mbox{\hspace{-0.33ex}} \tilde{S}^{(2 2)}_\tau (Q) \nonumber \\ 
& = & - \pi \, \epsilon (q_0) \left( \frac{1}{2} - n_F (|q_0|) 
\right) \, \displaystyle{ \raisebox{0.9ex}{\scriptsize{$\diamond$}}} 
\mbox{\hspace{-0.33ex}} \rho_\tau (Q) \, , 
\label{ImS11} \\ 
\displaystyle{ \raisebox{1.1ex}{\scriptsize{$\diamond$}}} 
\mbox{\hspace{-0.33ex}} \tilde{S}^{(1 2) / (2 1)}_\tau (Q) & = & \pm 
i \pi \, n_F (\pm q_0) \, \displaystyle{ 
\raisebox{0.9ex}{\scriptsize{$\diamond$}}} \mbox{\hspace{-0.33ex}} 
\rho_\tau (Q) \, . 
\label{S12} 
\end{eqnarray} 
Here $\epsilon (q_0) \equiv q_0 / |q_0|$, $n_F (x)$ $\equiv$ $1 / 
(e^{x / T} + 1)$, and 
\begin{equation} 
\displaystyle{ \raisebox{0.9ex}{\scriptsize{$\diamond$}}} 
\mbox{\hspace{-0.33ex}} \rho_\tau (Q) = \frac{\epsilon (q_0)}{\pi} 
\, Im \, \frac{1}{\displaystyle{ 
\raisebox{1.1ex}{\scriptsize{$\diamond$}}} \mbox{\hspace{-0.33ex}} 
D_\tau (Q)} \, . 
\label{dia-rho} 
\end{equation} 

The diagonalized FP-ghost propagator $\displaystyle{ 
\raisebox{1.1ex}{\scriptsize{$\diamond$}}} \mbox{\hspace{-0.33ex}} 
\Delta_F^g (Q)$ obtained from (\ref{qf-FP}) is 
\begin{equation} 
\displaystyle{ \raisebox{1.1ex}{\scriptsize{$\diamond$}}} 
\mbox{\hspace{-0.33ex}} \Delta_F^g (Q) = \frac{1}{Q^2 - \Pi_F^g (Q)} 
\, . 
\label{FP-prop} 
\end{equation} 
Using (\ref{qf-gluon}) and (\ref{2-1}) - (\ref{d}), we obtain 
\cite{lan,kobes} for the diagonalized gluon propagator, 
$\displaystyle{ \raisebox{1.1ex}{\scriptsize{$\diamond$}}} 
\mbox{\hspace{-0.33ex}} \Delta_F^{\mu \nu} (Q)$: 
\begin{eqnarray} 
\displaystyle{ \raisebox{1.1ex}{\scriptsize{$\diamond$}}} 
\mbox{\hspace{-0.33ex}} \Delta_F^{\mu \nu} (Q) & = & - 
{\cal P}_T^{\mu \nu} (Q) \, \displaystyle{ 
\raisebox{1.1ex}{\scriptsize{$\diamond$}}} \mbox{\hspace{-0.33ex}} 
\Delta^{T}_F (Q) - {\cal P}_L^{\mu \nu} (Q) \, \displaystyle{ 
\raisebox{1.1ex}{\scriptsize{$\diamond$}}} \mbox{\hspace{-0.33ex}} 
\Delta^{L}_F (Q) \nonumber \\ 
& & - {\cal C}^{\mu \nu} (Q) \, 
\displaystyle{ \raisebox{1.1ex}{\scriptsize{$\diamond$}}} 
\mbox{\hspace{-0.33ex}} \Delta^{C}_F (Q) - {\cal D}^{\mu \nu} (Q) \, 
\displaystyle{ \raisebox{1.1ex}{\scriptsize{$\diamond$}}} 
\mbox{\hspace{-0.33ex}} \Delta^{D}_F (Q) \, , 
\label{g-p-sedu} \\ 
\displaystyle{ \raisebox{1.1ex}{\scriptsize{$\diamond$}}} 
\mbox{\hspace{-0.33ex}} \Delta^{T}_F (Q) & = & \frac{1}{Q^2 - 
\Pi_F^T (Q)} \, , 
\label{g-p-t} \\ 
\displaystyle{ \raisebox{1.1ex}{\scriptsize{$\diamond$}}} 
\mbox{\hspace{-0.33ex}} \Delta^{L}_F (Q) & = & \frac{1}{Q^2 - 
\Pi_F^L (Q)} \, , 
\label{g-p-l} \\ 
\displaystyle{ \raisebox{1.1ex}{\scriptsize{$\diamond$}}} 
\mbox{\hspace{-0.33ex}} \Delta^{C}_F (Q) & = & \eta \, 
\frac{\Pi_F^C (Q)}{(Q^2 - \Pi_F^L (Q)) (Q^2 + i 0^+)} \, , 
\label{g-p-c} \\ 
\displaystyle{ \raisebox{1.1ex}{\scriptsize{$\diamond$}}} 
\mbox{\hspace{-0.33ex}} \Delta^{D}_F (Q) & = & \eta \, \frac{1}{Q^2 
+ i 0^+} \, , 
\label{g-p-d} 
\end{eqnarray} 

\noindent where use has been made of (\ref{piD0}). $2 \times 2$ 
FP-ghost- and gluon-propagators are obtained, respectively, from 
$\displaystyle{ \raisebox{1.1ex}{\scriptsize{$\diamond$}}} 
\mbox{\hspace{-0.33ex}} \Delta_F^g (Q)$ and  
$\displaystyle{ \raisebox{1.1ex}{\scriptsize{$\diamond$}}} 
\mbox{\hspace{-0.33ex}} \Delta_F^{\mu \nu} (Q)$ through Bogoliubov 
transformation \cite{lan}: 
\begin{eqnarray} 
\displaystyle{ \raisebox{1.1ex}{\scriptsize{$\diamond$}}} 
\mbox{\hspace{-0.33ex}} \Delta^{g / \mu \nu \, (1 1)} (Q) & = & - 
\left[ \displaystyle{ \raisebox{1.1ex}{\scriptsize{$\diamond$}}} 
\mbox{\hspace{-0.33ex}} \Delta^{g / \mu \nu \, (2 2)} (Q) \right]^* 
\nonumber \\ 
& = & 
[1 + n_B (|q_0|)] \, \displaystyle{ 
\raisebox{1.1ex}{\scriptsize{$\diamond$}}} \mbox{\hspace{-0.33ex}} 
\Delta_F^{g / \mu \nu} (Q) - n_B (|q_0|) \left( \displaystyle{ 
\raisebox{1.1ex}{\scriptsize{$\diamond$}}} \mbox{\hspace{-0.33ex}} 
\Delta_F^{g / \mu \nu} (Q) \right)^* \, , 
\label{mat-1} \\ 
\displaystyle{ \raisebox{1.1ex}{\scriptsize{$\diamond$}}} 
\mbox{\hspace{-0.33ex}} \Delta^{g / \mu \nu \, (1 2)} (Q) & = & 
[\theta (- q_0) + n_B (|q_0|)] \left[ \displaystyle{ 
\raisebox{1.1ex}{\scriptsize{$\diamond$}}} \mbox{\hspace{-0.33ex}} 
\Delta_F^{g / \mu \nu} (Q) - \left( \displaystyle{ 
\raisebox{1.1ex}{\scriptsize{$\diamond$}}} \mbox{\hspace{-0.33ex}} 
\Delta_F^{g / \mu \nu} (Q) \right)^* \right] \, , 
\label{mat-2} \\ 
\displaystyle{ \raisebox{1.1ex}{\scriptsize{$\diamond$}}} 
\mbox{\hspace{-0.33ex}} \Delta^{g / \mu \nu \, (2 1)} (Q) & = & 
[\theta (q_0) + n_B (|q_0|)] \left[ \displaystyle{ 
\raisebox{1.1ex}{\scriptsize{$\diamond$}}} \mbox{\hspace{-0.33ex}} 
\Delta_F^{g / \mu \nu} (Q) - \left( \displaystyle{ 
\raisebox{1.1ex}{\scriptsize{$\diamond$}}} \mbox{\hspace{-0.33ex}} 
\Delta_F^{g / \mu \nu} (Q) \right)^* \right] \, . 
\label{mat-3} 
\end{eqnarray} 

Here we recall that the invariance of 
${\cal L}_{\mbox{\scriptsize{QCD}}}$ under the BRS transformation 
leads to the Ward-Takahashi relation \cite{lan}, 
\begin{eqnarray}
Q^\nu \, \Delta^{' a b \, (r s)}_{\mu \nu} (Q) & = & - \eta \left[ 
Q_\mu \, \Delta^{' a b \, (r s)}_g (Q) \right. \nonumber \\ 
& & \left. - (-)^{r - 1} \Pi^{' a c \, (r t)}_{g; \, \mu} (Q) \, 
\Delta^{' c b \, (t s)}_g (Q) \right] \, , 
\label{BRS} 
\end{eqnarray}
where $r, s, t$ are thermal indexes, $a, b, c$ are color indexes, 
and $\Delta^{' a b}$s ($\equiv \delta^{a b} \, \Delta'$s) are full 
propagators. $\Pi^{' a c \, (r t)}_{g; \, \mu} (Q)$ is the 
pre-self-energy-part for a FP-ghost, which satisfies $Q^\mu \, 
\Pi^{' a c \, (r t)}_{g; \, \mu} (Q)$ $=$ $\Pi^{' a c \, 
(r t)}_g (Q)$. 

Multiplying $Q^\mu$ to the both sides of (\ref{BRS}) and using the 
Schwinger-Dyson equation for $\Delta^{' a b \, (r s)}_g (Q)$, one 
obtains \cite{lan} 
\begin{equation} 
\Delta^{' D \, (r s)} (Q) = \eta \, \Delta^{(0) \, (r s)}_g (Q) \, , 
\label{WT-1} 
\end{equation} 
where $\Delta_g^{(0) \, (r s)} (Q)$ is the bare FP-ghost propagator. 
Eq. (\ref{WT-1}) leads to (\ref{piD0}). 

Projecting out $\Delta^{' C \, (r s)} (Q)$, which is defined as in 
(\ref{g-p-sedu}), from (\ref{BRS}), we obtain 
\begin{eqnarray*} 
\Delta^{' C \, (r s)} (Q) & = & \eta \, \frac{\sqrt{2} q}{q_0} 
\left[ \Delta^{' (r s)}_g  (Q) - \Delta_g^{(0) \, (r s)} (Q) \right. 
\\ 
& & \left. + (-)^{r - 1} \frac{1}{q^2} \, \tilde{Q}_\mu \, \Pi^{' a 
b \, (r t)}_{g; \, \mu} (Q) \, \Delta^{' b a \, (t s)}_g (Q) \right] 
\, , 
\end{eqnarray*} 
where the sum is not taken over $a$ in the last term, which is 
independent of $a$, thanks to the $S U (N)$ symmetry. As will be 
seen in Sec. V, the region of our interest is $|Q^2|$ $<<$ $q^2$ $=$ 
$O (T^2)$, where the first and second terms in the square brackets 
are of $O (1 / Q^2)$, while the third term is of $O (1)$. Neglecting 
the third term, we obtain for the leading contribution to the 
diagonalized propagator, $\displaystyle{ 
\raisebox{1.1ex}{\scriptsize{$\diamond$}}} \mbox{\hspace{-0.33ex}} 
\Delta_F^C (Q)$, 
\begin{equation} 
\displaystyle{ \raisebox{1.1ex}{\scriptsize{$\diamond$}}} 
\mbox{\hspace{-0.33ex}} \Delta^C_F (Q) \simeq \eta \, 
\frac{\sqrt{2} q}{q_0} \, \frac{\Pi_F^g (Q)}{(Q^2 - \Pi_F^g (Q)) 
(Q^2 + i 0^+)} \, , 
\label{WT-3} 
\end{equation} 
where $|q_0| \simeq q$. Comparison of (\ref{g-p-c}) with 
(\ref{WT-3}) yields 
\begin{eqnarray} 
\Pi_F^L (Q) & \simeq & \Pi_F^g (Q) \, , 
\label{WT-4} \\ 
\Pi_F^C (Q) & \simeq & \sqrt{2} \, \epsilon (q_0) \, \Pi_F^L (Q) 
\, , 
\label{WT-5} 
\end{eqnarray} 
which are valid at $|Q^2| << q^2$. We choose $\Pi_F$s in 
(\ref{qf-gluon}) and (\ref{qf-FP}) so as to satisfy the relations 
(\ref{piD0}), (\ref{WT-4}), and (\ref{WT-5}). 
\setcounter{equation}{0}
\setcounter{section}{2}
\section{Hard-quark mode} 
In this section, we determine $\Sigma_F$ in (\ref{qf-quark}) 
self-con\-sist\-ent\-ly to one loop-order. The diagram to be 
analyzed is depicted in Fig. 1. In Fig. 1(a), for soft $K$ [$Q - 
K$], the HTL-resummed effective gluon [quark] propagator 
$\displaystyle{\raisebox{0.6ex}{\scriptsize{*}}} \! \Delta^{(j i)} 
(K)$ [$\displaystyle{\raisebox{0.6ex}{\scriptsize{*}}} \! S^{(j i)} 
(Q - K)$] should be assigned. It is to be noted that in calculating 
$\displaystyle{\raisebox{0.6ex}{\scriptsize{*}}} \! \Delta^{(j i)} 
(K)$ and $\displaystyle{\raisebox{0.6ex}{\scriptsize{*}}} \! 
S^{(j i)} (Q - K)$ the forms $\displaystyle{ 
\raisebox{1.1ex}{\scriptsize{$\diamond$}}}\mbox{\hspace{-0.33ex}} 
S^{(j i)}$ and $\displaystyle{ 
\raisebox{1.1ex}{\scriptsize{$\diamond$}}}\mbox{\hspace{-0.33ex}} 
\Delta^{(j i)}$, obtained in Sec. II, should be used for, 
respectively, hard-quark- and hard-gluon-propagators in the HTL. 

Let $\tilde{\Sigma}_F^{(1 a)}$ be the contribution of Fig. 1(a). A 
dimensional analysis shows that, for $|Q^2|$ $>>$ $g^2 T^2$, 
$|\tilde{\Sigma}_F^{(1 a)} (Q)|$ $<<$ $|Q^2| / q$ and, up to a 
possible factor of $\ln (g^{- 1})$, $\tilde{\Sigma}_F^{(1 a)}$ $=$ 
$O (g^2 T)$ for $|Q^2|$ $\leq$ $O (g^2 T^2)$. Then it is sufficient 
to analyze Fig. 1 in the region 
\begin{equation} 
||q_0| - q| \leq O (g^2 T) \, . 
\label{region} 
\end{equation} 

The contribution from Fig. 1(b), $\tilde{\Sigma}_F^{(1 b)}$, is 
obtained from (\ref{qf-quark}) and (\ref{int-1}): 
\begin{equation} 
\tilde{\Sigma}_F^{(1 b)} (Q) = - \Sigma_F (Q) \, . 
\label{1b} 
\end{equation} 

Computation of Fig. 1(a) in conventional hot QCD is carried out e.g. 
in \cite{honnin}, where it has been shown that, to leading order, 
\begin{equation} 
Re \, \tilde{\Sigma}_F^{(1 a)} (Q) \rule[-4mm]{0.2mm}{9mm} 
\raisebox{-3.4mm}{\scriptsize{\hspace{0.5mm}conventional}} \simeq 
m_f^2 \, \frac{q_0}{q^2} \, \gamma^0 \, . 
\label{conv-re-sigma} 
\end{equation} 
Here $m_f^2$ $=$ $g^2 C_F T^2 / 8$ with $C_F$ $=$ $(N^2 - 1) / 
(2 N)$. The leading contribution (\ref{conv-re-sigma}), being gauge 
independent, comes from the region where $K$ {\em and} $Q - K$ in 
Fig. 1(a) are hard. In other ward, the result (\ref{conv-re-sigma}) 
is insensitive to the soft-$K$ and soft-$(Q - K)$ region in Fig 
1(a). The difference between the thermal propagator $\displaystyle{ 
\raisebox{1.1ex}{\scriptsize{$\diamond$}}}\mbox{\hspace{-0.33ex}} 
S^{(j i)} (Q)$ $[\displaystyle{ 
\raisebox{1.1ex}{\scriptsize{$\diamond$}}}\mbox{\hspace{-0.33ex}} 
\Delta^{(j i)} (Q)]$ constructed in Sec. II and the one $S^{(j i)} 
(Q)$ $[\Delta^{(j i)} (Q)]$ in conventional hot QCD cannot be 
ignored at the region $|Q^2|$ $\leq$ $O (g^2 T^2)$. The result 
(\ref{conv-re-sigma}) is however insensitive to the region $|K^2|$, 
$|(Q - K)^2|$ $\leq$ $O (g^2 T^2)$, i.e., Fig. 1(a) with 
$\displaystyle{ 
\raisebox{1.1ex}{\scriptsize{$\diamond$}}}\mbox{\hspace{-0.33ex}} 
S^{(j i)} (Q - K)$ and $\displaystyle{ 
\raisebox{1.1ex}{\scriptsize{$\diamond$}}}\mbox{\hspace{-0.33ex}} 
\Delta^{(j i)} (K)$ for, respectively, the bare hard-quark and 
hard-gluon propagators yields, to leading order, the same result 
(\ref{conv-re-sigma}): 
\begin{equation} 
Re \, \tilde{\Sigma}_F^{(1 a)} (Q) \simeq m_f^2 \, \frac{q_0}{q^2} 
\, \gamma^0 \, . 
\label{re-sigma} 
\end{equation} 

Now we impose the self-consistency condition, 
\[ 
\tilde{\Sigma}_F (Q) = \tilde{\Sigma}_F^{(1 a)} (Q) + 
\tilde{\Sigma}^{(1 b)}_F (Q) = 0 \, , 
\] 
or 
\begin{equation} 
\Sigma_F (Q) = \tilde{\Sigma}_F^{(1 a)} (Q) \, , 
\label{self-consistency1} 
\end{equation} 
where use has been made of (\ref{1b}). Substituting (\ref{re-sigma}) 
into (\ref{self-consistency1}), we obtain 
\[ 
Re \, \Sigma_F (Q) \simeq m_f^2 \frac{q_0}{q^2} \, \gamma^0 
\] 
or, from (\ref{fg-def}), 
\begin{eqnarray} 
Re \, f (Q) & \simeq & 0 \, , \nonumber \\ 
Re \, g (Q) & \simeq & m_f^2 \, \frac{q_0}{q^2} \, . 
\label{3-30} 
\end{eqnarray} 
Here $Re \, f (Q) \simeq 0$ means $|Re \, f (Q)|$ $<<$ $g^2 T$. 

We are now in a position to compute $Im \, 
\tilde{\Sigma}_F^{(1 a)}$. Shown in \cite{honnin} is that, in 
conventional hot QCD calculation, the leading contribution (at 
logarithmic accuracy) to $Im \, 
\tilde{\Sigma}_F^{(1 a)} \rule[-4mm]{0.2mm}{9mm} 
\raisebox{-3.4mm}{\scriptsize{\hspace{0.5mm}conventional}}$ comes 
from Fig. 1(a) with soft $K$. More precisely, the magnetic part of 
the soft-gluon propagator yields the leading contribution. In other 
words, the contributions from the electric part and from the 
gauge-parameter-dependent part of the soft-gluon propagator are 
nonleading. This means in particular that the leading contribution 
is gauge independent. In contrast to the case of $Re \, 
\tilde{\Sigma}_F^{(1 a)}$, $Im \, \tilde{\Sigma}_F^{(1 a)}$ is 
logarithmically sensitive to the region $(Q - K)^2$ $\simeq$ $0$, so 
that we should compute $Im \, \tilde{\Sigma}_F^{(1 a)}$ in the 
theory defined in Sec II. 

As has been mentioned above, we are interested in the region 
(\ref{region}). In this region, $|q_0|$ $(\equiv \tau q_0)$ $\simeq$ 
$q$, we write $f (Q)$ and $g (Q)$ in (\ref{fg-def}) 
\begin{equation} 
f_\tau (q_0, q) \equiv f(Q) \, , \;\;\;\;\;\;\; g_\tau (q_0, q) 
\equiv g (Q) \, , 
\label{3-31} 
\end{equation} 
which satisfies $f_\tau (q_0, q)$ $=$ $f_{- \tau} (- q_0, q)$ and 
$g_\tau (q_0, q)$ $=$ $- g_{- \tau} (- q_0, q)$.  

Leading contribution to $Im \, \tilde{\Sigma}_F^{(1 a)}$ reads 
\cite{lan}, 
\begin{eqnarray} 
Im \, \tilde{\Sigma}_F^{(1 a)} (Q) & = & \frac{i}{2 [\theta (q_0) - 
n_F (q)]} \tilde{\Sigma}^{(2 1)}_{(1 a)} (Q) \nonumber \\ 
& \simeq & \frac{g^2 \, C_F}{2 [\theta (q_0) - n_F (q)]} 
\int_{\mbox{\scriptsize soft} \, K} \frac{d^{\, 4} K}{(2 \pi)^4} \, 
\gamma_\rho \nonumber \\ & & \times \displaystyle{ 
\raisebox{1.1ex}{\scriptsize{$\diamond$}}} \mbox{\hspace{-0.33ex}} 
S^{(2 1)} (Q - K) \, \gamma_\sigma \, 
\displaystyle{\raisebox{0.6ex}{\scriptsize{*}}} \! \Delta_{\rho 
\sigma}^{\mbox{\scriptsize{mag}} (2 1)} (K) \, , 
\label{Im-start} 
\end{eqnarray} 
where $\displaystyle{\raisebox{0.6ex}{\scriptsize{*}}} \! 
\Delta_{\rho \sigma}^{\mbox{\scriptsize{mag}} (2 1)}$ is the $(2, 
1)$-component of the magnetic part of the effective gluon propagator 
\cite{gluon}, 
\begin{eqnarray} 
\displaystyle{\raisebox{0.6ex}{\scriptsize{*}}} \! 
\Delta_{\rho \sigma}^{\mbox{\scriptsize{mag}} (2 1)} (K) & \equiv & 
- {\cal P}_{T \, \rho \sigma} (K) \, 
\displaystyle{\raisebox{0.6ex}{\scriptsize{*}}} \! 
\tilde{\Delta}^{\mbox{\scriptsize{mag}} (2 1)} (K) \, , 
\label{delta-star} \\ 
\displaystyle{\raisebox{0.6ex}{\scriptsize{*}}} \! 
\tilde{\Delta}^{\mbox{\scriptsize{mag}} (2 1)} (K) & \simeq & 
\displaystyle{\raisebox{0.6ex}{\scriptsize{*}}} \! 
\tilde{\Delta}^{\mbox{\scriptsize{mag}} (i j)} (K) \simeq - 2 \pi i 
\, \frac{T}{k_0} \, \rho_t (K) \, , \nonumber \\ 
& & \mbox{\hspace*{20ex}} (i, j = 1, 2) \, . 
\label{til-delta-star} 
\end{eqnarray} 

\noindent Here ${\cal P}_{T \, \rho \sigma} (K)$ is as in (\ref{pt}) 
and $\rho_t (K)$ is the spectral function. As has been mentioned at 
the beginning of this section, in computing 
$\displaystyle{\raisebox{0.6ex}{\scriptsize{*}}} \! \Delta_{\rho 
\sigma}^{\mbox{\scriptsize{mag}} (2 1)} (K)$, we should use 
$\displaystyle{ 
\raisebox{1.1ex}{\scriptsize{$\diamond$}}}\mbox{\hspace{-0.33ex}} 
S^{(j i)}$ and $\displaystyle{ 
\raisebox{1.1ex}{\scriptsize{$\diamond$}}}\mbox{\hspace{-0.33ex}} 
\Delta^{(j i)}$ for hard propagators in the HTL. It is not difficult 
to see that the result (\ref{Im-sigma-15}) below is not sensitive to 
this \lq\lq modification'' of the hard propagators. Then, in the 
following, we shall use the form of 
$\displaystyle{\raisebox{0.6ex}{\scriptsize{*}}} \! 
\Delta_{\rho \sigma}^{\mbox{\scriptsize{mag}} (2 1)} (K)$, computed 
in conventional hot QCD \cite{gluon}. 

Substituting (\ref{q-pro-1}), (\ref{delta-star}), and 
(\ref{til-delta-star}) into (\ref{Im-start}), 
we obtain 
\begin{eqnarray} 
Im \, \tilde{\Sigma}_F^{(1 a)} (Q) & \simeq & - \frac{i g^2 \, C_F 
\, T}{\theta (q_0) - n_F (q)} \int_{\mbox{\scriptsize{soft}} \, K} 
\frac{d^{\, 4} K}{(2 \pi)^3} \, \frac{1}{k_0} \, \rho_t (K) 
\nonumber \\ 
& & \times \sum_{\tau = \pm} \left[ 
\hat{{Q\kern-0.1em\raise0.3ex\llap{/}\kern0.15em\relax}}_\tau + \tau 
\hat{{\bf q}} \cdot \vec{\gamma} - \tau (\vec{\gamma} \cdot 
\hat{{\bf k}}) (\hat{{\bf k}} \cdot \hat{{\bf q}}) \right] \nonumber 
\\ 
& & \times \displaystyle{ \raisebox{1.1ex}{\scriptsize{$\diamond$}}} 
\mbox{\hspace{-0.33ex}} \tilde{S}_\tau^{(2 1)} (Q - K) \, . 
\label{Im-sigma-3} 
\end{eqnarray} 
$\displaystyle{ \raisebox{1.1ex}{\scriptsize{$\diamond$}}} 
\mbox{\hspace{-0.33ex}} \tilde{S}_\tau^{(2 1)} (P)$ with $P$ $=$ $Q 
- K$ is obtained from (\ref{S12}) with (\ref{dia-rho}), 
(\ref{dia-D}), and (\ref{3-31}): 
\begin{eqnarray*} 
\displaystyle{ \raisebox{1.1ex}{\scriptsize{$\diamond$}}} 
\mbox{\hspace{-0.33ex}} \tilde{S}_\tau^{(2 1)} (P) & \simeq & 
[\theta (p_0) - n_F (p)] \\ 
& & \times \frac{g_\tau^{(i)} + (p_0 - \tau p) f_\tau^{(i)}}{[(p_0 
- \tau p) \{ 1 - f_\tau^{(r)} \} - g_\tau^{(r)} ]^2 + [g_\tau^{(i)} 
+ (p_0 - \tau p) f_\tau^{(i)}]^2} \, , 
\end{eqnarray*} 
where $f_\tau^{(r)} \equiv Re \, f_\tau^{(r)} (p_0, p)$, 
$f_\tau^{(i)}$ $=$ $Im \, f_\tau^{(i)} (p_0, p)$, etc. Using 
(\ref{3-30}) with (\ref{3-31}), we have 
\begin{equation} 
\displaystyle{ \raisebox{1.1ex}{\scriptsize{$\diamond$}}} 
\mbox{\hspace{-0.33ex}} \tilde{S}_\tau^{(2 1)} (Q - K) \simeq i \, 
[\theta (q_0) - n_F (q)] \, \frac{\tilde{g}_\tau^{(i)} 
(q_0, q)}{[q_0 - \tau (q + m_f^2 / q) - k_0 + \tau {\bf k} \cdot 
\hat{{\bf q}}]^2 + [\tilde{g}_\tau^{(i)} (q_0, q) ]^2} \, , 
\label{dia-S12-3} 
\end{equation} 
where $Q$ is hard, $K$ is soft, and 
\begin{equation} 
\tilde{g}_\tau^{(i)} (q_0, q) \equiv g_\tau^{(i)} (q_0, q) + (q_0 - 
\tau q) f_\tau^{(i)} (q_0, q) \, . 
\label{yametai} 
\end{equation} 
We shall show below (cf. (\ref{pre-final}) with (\ref{fg-def})) that 
the second term on the R.H.S. of (\ref{yametai}) turns out to be 
negligible when compared to the first term, so that 
\[ 
\tilde{g}_\tau^{(i)} (q_0, q) \simeq g_\tau^{(i)} (q_0, q) \, . 
\] 

Substituting (\ref{dia-S12-3}) into (\ref{Im-sigma-3}), we obtain 
\begin{eqnarray} 
Im \, \tilde{\Sigma}_F^{(1 a)} (Q) & \simeq & g^2 \, C_F \, T \, 
\int_{\mbox{\scriptsize{soft}} \, K} \frac{d^{\, 4} K}{(2 \pi)^3} \, 
\frac{\rho_t (K)}{k_0} \left[ 
\hat{{Q\kern-0.1em\raise0.3ex\llap{/}\kern0.15em\relax}}_\tau + \tau 
\hat{{\bf q}} \cdot \vec{\gamma} - \tau (\vec{\gamma} \cdot 
\hat{{\bf k}}) (\hat{{\bf k}} \cdot \hat{{\bf q}}) \right] \nonumber 
\\ 
& & \times \frac{g_\tau^{(i)} (q_0, q)}{[q_0 - \tau (q + m_f^2 / q) 
- k_0 + \tau {\bf k} \cdot \hat{{\bf q}}]^2 + [g_\tau^{(i)} (q_0, q) 
]^2} \, . 
\label{Im-sigma-10} 
\end{eqnarray} 

It is well known \cite{damp,smilga,pis5} that, at logarithmic 
accuracy, the dominant contribution comes from the region where 
$|k_0|$ $<<$ $k$ and $|\hat{{\bf k}} \cdot \hat{{\bf q}}|$ $<<$ $1$. 
Then, the piece $- \tau (\vec{\gamma} \cdot \hat{{\bf k}}) 
(\hat{{\bf k}} \cdot \hat{{\bf q}})$ in (\ref{Im-sigma-10}) leads to 
a nonleading contribution. Setting $k_0 = 0$ in the denominator of 
the last term in (\ref{Im-sigma-10}) and integrating over $-\alpha$ 
$<$ $\hat{{\bf k}} \cdot \hat{{\bf q}}$ $<$ $+ \alpha$ with $\alpha 
<< 1$, we obtain 
\begin{eqnarray} 
Im \, \tilde{\Sigma}_F^{(1 a)} (Q) & \simeq & \frac{g^2}{4 \pi^2} \, 
C_F \, T \, \gamma^0 \int_{\mbox{\scriptsize{soft}} \, K} d k \, k 
\int \frac{d k_0}{k_0} \, \rho_t (K) \, \frac{g_i}{|g_i|}  \nonumber 
\\ 
& & \times \sum_{\xi = \pm} \arctan \left( \frac{\alpha k + \xi \, 
\{ q_0 - \tau (q + m_f^2 / q) \}}{|g_\tau^{(i)} (q_0, q)|} \right) 
\, . 
\label{Im-sigma-11} 
\end{eqnarray} 
As mentioned at the beginning of this section, up to a 
possible factor of $\ln (g^{- 1})$, $|g_\tau^{(i)} (q_0, q)|$ is of 
$O (g^2 T)$ and we are interested in the region $|q_0 - \tau (q + 
m_f^2 / q)|$ $=$ $O (g^2 T)$ (cf. (\ref{region})). At the region 
$|k_0|$ $<<$ $k$, 
\begin{equation} 
\rho_t (K) \simeq {\cal M}_T^2 \, \frac{k \, k_0}{k^6 + (\pi 
{\cal M}_T^2)^2 \, k_0^2} \, , 
\label{small-k0}
\end{equation} 
where ${\cal M}_T^2 \equiv 3 m_T^2 / 4$ with 
\begin{equation} 
m_T^2 = \frac{1}{9} \left( N + \frac{N_f}{2} \right) (g T)^2 \, . 
\label{gluon-mass} 
\end{equation} 
Substituting (\ref{small-k0}) into (\ref{Im-sigma-11}) and 
integrating over $- \beta k$ $<$ $k_0$ $<$ $+ \beta k$ with $\beta$ 
$<<$ $1$, we get 
\begin{eqnarray*} 
Im \, \tilde{\Sigma}_F^{(1 a)} (Q) & \simeq & \frac{2}{\pi^2} \, 
\frac{g^2}{4 \pi} \, C_F \, T \, \gamma^0 \, 
\int_{\mbox{\scriptsize{soft}} \, k} \frac{d k}{k} \, \arctan \left( 
\frac{\pi \beta {\cal M}_T^2}{k^2} \right) \, \frac{g_\tau^{(i)} 
(q_0, q)}{|g_\tau^{(i)} (q_0, q)|}  
\nonumber \\ 
& & \times \sum_{\xi = \pm} \arctan \left( \frac{\alpha k + \xi \, 
\{ q_0 - \tau (q + m_f^2 / q) \}}{|g_\tau^{(i)} (q_0, q)|} \right) 
\, . 
\end{eqnarray*} 
Now we observe that $\arctan \{\pi \beta {\cal M}_T^2 / 
k^2 \}$ $=$ $\pi / 2$ at $k$ $=$ $0$ and $\propto$ $\beta m_T^2 / 
k^2$ $\propto$ $(g T / k)^2$ for $k$ $>>$ $g T$. The transition 
region is $k$ $=$ $O (g T)$. The quantity at the second line 
vanishes at $k$ $=$ $0$ and $\simeq \pi$ for $k$ $>>$ 
$|g_\tau^{(i)}|$. When $O \{|q_0 - \tau (q + m_f^2 / q)| \}$ $\leq$ 
$O (|g_\tau^{(i)}|)$, the transition region is $k$ $=$ $O 
(|g_\tau^{(i)}|)$ and, when $O \{|q_0 - \tau (q + m_f^2 / q)| \}$ 
$>$ $O (|g_\tau^{(i)}|)$, the transition region is $k$ $=$ $O \{|q_0 
- \tau (q + m_f^2 / q)| \}$. 

>From the above observation, we obtain for the leading contribution 
at logarithmic accuracy, 
\begin{eqnarray} 
Im \, \tilde{\Sigma}_F^{(1 a)} (Q) & \simeq & \frac{g^2}{4 \pi} \, 
C_F \, T \, \gamma^0 \, \frac{g_\tau^{(i)} (q_0, q)}{|g_\tau^{(i)} 
(q_0, q)|} \ln \left( \frac{m_T}{\Gamma_q (Q)} \right) \, , 
\label{Im-sigma-15} 
\end{eqnarray} 
where 
\[ 
\Gamma_q (Q) \equiv max \left[ |q_0 - \tau (q + m_f^2 / q)| , \, 
|g_\tau^{(i)} (q_0, q)| \right] \, . 
\] 
It should be emphasized again that (\ref{Im-sigma-15}) is valid at 
logarithmic accuracy, i.e., the term of $O (1)$ is ignored when 
compared to $\ln \{ m_T / \Gamma_q (Q) \}$. 

Substituting (\ref{Im-sigma-15}) into the self-consistency condition 
(\ref{self-consistency1}), we obtain 
\begin{equation} 
Im \, \Sigma_F (Q) \simeq \frac{g^2}{4 \pi} \, C_F \, T \, \gamma^0 
\, \frac{g_\tau^{(i)} (q_0, q)}{|g_\tau^{(i)} (q_0, q)|} \ln \left( 
\frac{m_T}{\Gamma_q (Q)} \right) \, . 
\label{pre-final} 
\end{equation} 
>From (\ref{fg-def}), (\ref{3-31}), and (\ref{pre-final}), we have 
\begin{eqnarray} 
f_\tau^{(i)} (q_0, q) & \simeq & 0 \, , 
\label{tsuika} \\ 
g_\tau^{(i)} (q_0, q) & = & \frac{g^2}{4 \pi} \, C_F \, T \, 
\frac{g_\tau^{(i)} (q_0, q)}{|g_\tau^{(i)} (q_0, q)|} \, \ln \left( 
\frac{m_T}{max \left[ |q_0 - \tau (q + m_f^2 / q)|, \, |g_\tau^{(i)} 
(q_0, q)| \right] } \right) \, . \nonumber \\ 
\label{kettei-eq} 
\end{eqnarray} 
It is to be noted that, in conventional hot QCD, we have 
\cite{honnin} (\ref{Im-sigma-15}) with \\ 
$g_\tau^{(i)} (q_0, q) / |g_\tau^{(i)} (q_0, q)|$ $=$ $- 
\epsilon(q_0)$ and $\Gamma_q (Q)$ $=$ $|q_0 - \tau (q + m_f^2 / 
q)|$. We also note that, for $q_0$ $>$ $0$, 
\begin{eqnarray} 
- i \, \epsilon (q_0) \, tr \left[ 
{Q\kern-0.1em\raise0.3ex\llap{/}\kern0.15em\relax} \, 
\tilde{\Sigma}_{(1 a)}^{(2 1)} (Q) \right] & = & - 2 \epsilon (q_0) 
[\theta (q_0) - n_F (q)] \nonumber \\ 
& & \times tr \left[ 
{Q\kern-0.1em\raise0.3ex\llap{/}\kern0.15em\relax} \, Im \, 
\tilde{\Sigma}_F^{(1 a)} (Q) \right] \, , 
\label{hikaku} 
\end{eqnarray} 
with $\tilde{\Sigma}^{(2 1)}_{(1 a)}$ as in (\ref{Im-start}), is 
proportional to the decay rate of a quark mode, whose propagator is 
given by (\ref{q-pro-1}) - (\ref{dia-rho}). While for $q_0$ $<$ $0$, 
(\ref{hikaku}) is proportional to the production rate. Then 
(\ref{hikaku}) should be positive, which means again that 
$g_\tau^{(i)} (q_0, q) / |g_\tau^{(i)} (q_0, q)|$ $=$ $ - \epsilon 
(q_0)$. From these observations, as the physically sensible 
solution, we assume that $g_\tau^{(i)} (q_0, q) / |g_\tau^{(i)} 
(q_0, q)|$ $=$ $ - \epsilon (q_0)$.  

As has been mentioned repeatedly, we see from (\ref{kettei-eq}) that 
the contribution of $O \{ g^2 T \ln (g^{- 1}) \}$ to $g^{(i)}_\tau$ 
emerges in the region, 
\begin{equation} 
||q_0| - q| \leq O \{g^2 T \ln (g^{- 1}) \} \, . 
\label{ryouiki} 
\end{equation} 
In fact, by taking the logarithm of (\ref{kettei-eq}), we can solve 
the resulting equation iteratively with respect to $\ln 
|g_\tau^{(i)} (q_0, q)|$ to obtain 
\begin{eqnarray} 
g_\tau^{(i)} (q_0, q) & \equiv & - \epsilon (q_0) \, \gamma_q 
\nonumber \\ 
\gamma_q & = & \frac{g^2}{4 \pi} \, C_F \, T \, \ln (g^{- 1}) \left[ 
1 - \frac{\ln \{ \ln (g^{- 1}) \}}{\ln (g^{- 1})} + {\cal F} \right] 
\nonumber \\ 
& & + O (g^2 T) \, . 
\label{quark-2} 
\end{eqnarray} 
When $||q_0| - q|$ $=$ $O \{ g^2 T \ln (g^{- 1}) \}$, ${\cal F}$ $=$ 
$0$, while for $||q_0| - q|$ $<$ $O \{ g^2 T \ln (g^{- 1}) \}$, 
${\cal G}$ $\equiv$ $1 - \ln \{ \ln (g^{- 1}) \} / \ln (g^{- 1})$ 
$+$ ${\cal F}$ is determined through ${\cal G}$ $=$ $1$ $-$ $\ln 
\{{\cal G} \ln (g^{- 1}) \} / \ln (g^{- 1})$. 

Here we summarize the results obtained above. From (\ref{fg-def}) 
with (\ref{3-31}), (\ref{3-30}), (\ref{tsuika}), and 
(\ref{quark-2}), we have for the self-consistently determined 
$\Sigma_F (Q)$, 
\[ 
\Sigma_F (Q) \simeq \tau \left[ \frac{m_f^2}{q} - i \gamma_q (Q) 
\right] \, \gamma^0 \, , 
\] 
where $\tau = \epsilon (q_0)$. It should be mentioned that we have 
evaluated $\gamma_q$ at logarithmic accuracy. Namely, the 
computation of the $O (g^2 T)$ contribution to $\gamma_q$, Eq. 
(\ref{quark-2}), is outside the scope of this paper. Taking this 
fact into account, we obtain from (\ref{dia-D}) with (\ref{3-31}), 
(\ref{3-30}), (\ref{tsuika}), and (\ref{quark-2}), 
\begin{eqnarray*} 
Re \, \frac{1}{\displaystyle{ 
\raisebox{1.1ex}{\scriptsize{$\diamond$}}} \mbox{\hspace{-0.33ex}} 
D_\tau (q_0, q)} & \simeq & - \frac{q_0 - \tau (q + m_f^2 / 
q) }{[q_0 - \tau (q + m_f^2 / q)]^2 + \gamma_q^2} \\ 
& \simeq & - \frac{q_0 - \tau (q + m_f^2 / q) }{(q_0 - \tau q)^2 + 
\gamma_q^2} \, , \\ 
Im \, \frac{1}{\displaystyle{ 
\raisebox{1.1ex}{\scriptsize{$\diamond$}}} \mbox{\hspace{-0.33ex}} 
D_\tau (q_0, q)} & \simeq & \frac{ \tau \, \gamma_q}{(q_0 - 
\tau q)^2 + \gamma_q^2} \, . 
\end{eqnarray*} 
Thus the resummation of the imaginary-part of the self-energy part, 
$Im \, \tilde{\Sigma}_F$, plays a dominant role. 

Then, the thermal propagator of the \lq\lq good'' mode with hard 
momentum $Q$ reads (cf. (\ref{q-pro-1}) - (\ref{dia-rho})) 
\begin{eqnarray} 
\displaystyle{ \raisebox{1.1ex}{\scriptsize{$\diamond$}}} 
\mbox{\hspace{-0.33ex}} S^{(j i)} (Q) & = & \sum_{\tau = \pm} 
\hat{{Q\kern-0.1em\raise0.3ex\llap{/}\kern0.15em\relax}}_\tau 
\displaystyle{ \raisebox{1.1ex}{\scriptsize{$\diamond$}}} 
\mbox{\hspace{-0.33ex}} \tilde{S}^{(j i)}_\tau (Q) \, , \;\;\;\; 
(j, \, i = 1, 2) \, , 
\label{kekka-1} \\ 
Re \, \displaystyle{ \raisebox{1.1ex}{\scriptsize{$\diamond$}}} 
\mbox{\hspace{-0.33ex}} \tilde{S}^{(1 1)}_\tau (Q) & = & - Re \, 
\displaystyle{ \raisebox{1.1ex}{\scriptsize{$\diamond$}}} 
\mbox{\hspace{-0.33ex}} \tilde{S}^{(2 2)}_\tau (Q) \nonumber \\ 
& \simeq & \frac{1}{2} \, \frac{q_0 - \epsilon (q_0) (q + m_f^2 / 
q)}{(q_0 - \tau q)^2 + \gamma_q^2} \, , 
\label{kekka-11} \\ 
Im \, \displaystyle{ \raisebox{1.1ex}{\scriptsize{$\diamond$}}} 
\mbox{\hspace{-0.33ex}} \tilde{S}^{(1 1)}_\tau (Q) & = & Im \, 
\displaystyle{ \raisebox{1.1ex}{\scriptsize{$\diamond$}}} 
\mbox{\hspace{-0.33ex}} \tilde{S}^{(2 2)}_\tau (Q) \nonumber \\ 
& \simeq & - \pi \, \epsilon (q_0) \left[ \frac{1}{2} - n_F (q) 
\right] \displaystyle{ \raisebox{0.9ex}{\scriptsize{$\diamond$}}} 
\mbox{\hspace{-0.33ex}} \rho_\tau (Q) \, , 
\label{kekka-2} \\ 
\displaystyle{ \raisebox{1.1ex}{\scriptsize{$\diamond$}}} 
\mbox{\hspace{-0.33ex}} \tilde{S}^{(1 2) / (2 1)}_\tau (Q) 
& \simeq & - i \pi \, \epsilon (q_0) [\theta (\mp q_0) - n_F (q)] \, 
\displaystyle{ \raisebox{0.9ex}{\scriptsize{$\diamond$}}} 
\mbox{\hspace{-0.33ex}} \rho_\tau (Q) \, , 
\label{kekka-3} 
\end{eqnarray} 
where 
\begin{equation} 
\displaystyle{ \raisebox{0.9ex}{\scriptsize{$\diamond$}}} 
\mbox{\hspace{-0.33ex}} \rho_\tau (Q) \simeq \frac{1}{\pi} \, 
\frac{\gamma_q}{(q_0 - \tau q)^2 + \gamma_q^2} \, . 
\label{kekka-4} 
\end{equation} 

The forms (\ref{kekka-2}) and (\ref{kekka-3}) are valid in the 
region (\ref{ryouiki}), while the form (\ref{kekka-11}) is valid in 
the region $O (g^3 T)$ $<$ $|q_0 - \epsilon (q_0) (q + m_f^2 / q)|$ 
$\leq$ $O \{ g^2 T \ln (g^{- 1}) \}$. For obtaining $Re \, 
\displaystyle{ \raisebox{1.1ex}{\scriptsize{$\diamond$}}} 
\mbox{\hspace{-0.33ex}} \tilde{S}^{(1 1)}_\tau (Q)$ in the region 
$|q_0 - \epsilon (q_0)  (q + m_f^2 / q)|$ $\leq$ $O (g^3 T)$, 
concrete evaluation of Fig. 1(a) as well as the two-loop 
contribution is necessary. 
\setcounter{equation}{0}
\setcounter{section}{3}
\section{Absence of additional contributions of leading order} 
\def\theequation{\mbox{\arabic{section}.\arabic{equation}}}
In this section, we analyze some other formally higher-order 
corrections to the hard-quark self-energy part and show that they 
are nonleading. 
\subsection{Analysis of Figs. 2 - 5} 
As has been recognized from the analysis in Sec. III, in 
conventional hot QCD, resummation of the one-loop self-energy part 
should be carried out for a thermal propagator of a hard quark close 
to the mass shell, $||q_0| - q|$ $ \leq $ $O \{ g^2 T \ln (g^{- 1}) 
\}$. It was shown, e.g., in \cite{honnin} that the same \lq\lq 
phenomenon'' occurs in the case of quark-gluon vertex. An one-loop 
contribution to the quark-gluon vertex is depicted in Fig. 2, where 
$K$ is soft and $Q$ is hard. When $Q$ and $Q - K$ are close to the 
mass shell, $|Q^2|$, $|(Q - K)^2|$ $\leq$ $O \{g^2 T^2 \ln (g^{- 1}) 
\}$, the contribution of Fig. 2 is of the same order of magnitude as 
the bare counterpart. As in the self-energy case, the leading 
contribution comes from the magnetic part of the soft-gluon 
propagator in Fig. 2 and thus is gauge independent. The same is true 
for multi-loop contributions. 

Fig. 2 yields 
\begin{eqnarray} 
\left( \Lambda^{a; \, \mu}_{g 1} (Q - K, Q) \right)^\ell_{j i} 
& \simeq & i g^2 (-)^{i + j + \ell} \left( \frac{N}{2} - C_F \right) 
\, T^a \nonumber \\ 
& & \times \int_{\mbox{\scriptsize{soft}} \, P} \frac{d^{\, 4} 
P}{(2 \pi)^4} \, \gamma^\sigma \, 
\raisebox{1.1ex}{\scriptsize{$\diamond$}} \mbox{\hspace{-0.33ex}} 
S^{(j \ell)} (Q - K + P) \, \gamma^\mu \nonumber \\ 
& & \times \raisebox{1.1ex}{\scriptsize{$\diamond$}} 
\mbox{\hspace{-0.33ex}} S^{(\ell i)} (Q + P) \, \gamma^\rho \, 
\displaystyle{\raisebox{0.6ex}{\scriptsize{*}}} \! \Delta_{\rho 
\sigma}^{\mbox{\scriptsize{mag}} (i j)} (P) \, , 
\label{g1-1} 
\end{eqnarray} 
where $T^a$ are the (hermitian) fundamental-representation matrix of 
$su(N)$. Substituting (\ref{delta-star}), (\ref{til-delta-star}), 
and (\ref{kekka-1}), we obtain, 
after some manipulation,  
\begin{eqnarray} 
\left( \Lambda^{a; \, \mu}_{g 1} (Q - K, Q) \right)^\ell_{j i} 
& \simeq & 4 g^2 (-)^{i + j + \ell} \left( \frac{N}{2} - C_F \right) 
T^a \, T \nonumber \\ 
& & \times \sum_{\tau = \pm} \hat{Q}_\tau 
\int_{\mbox{\scriptsize{soft}} \, P} \frac{d^{\, 4} P}{(2 \pi)^3} \, 
\left[ \hat{{Q\kern-0.1em\raise0.3ex\llap{/}\kern0.15em\relax}}_\tau 
+ \tau \hat{{\bf q}} \cdot \vec{\gamma} - \tau (\vec{\gamma} \cdot 
\hat{{\bf p}}) (\hat{{\bf p}} \cdot \hat{{\bf q}}) \right] \nonumber 
\\ 
& & \times \frac{\rho_t (P)}{p_0} 
\raisebox{1.1ex}{\scriptsize{$\diamond$}} \mbox{\hspace{-0.33ex}} 
\tilde{S}^{(j \ell)}_\tau (Q - K + P) \, 
\raisebox{1.1ex}{\scriptsize{$\diamond$}} \mbox{\hspace{-0.33ex}} 
\tilde{S}_\tau^{(\ell i)} (Q + P) \, . \nonumber 
\end{eqnarray} 
As will be shown in Appendix A, the dominant contribution 
comes from the region where $|\hat{{\bf p}} \cdot \hat{{\bf q}}|$ 
$<<$ $1$, so that 
\begin{eqnarray*} 
& & \hat{{Q\kern-0.1em\raise0.3ex\llap{/}\kern0.15em\relax}}_\tau + 
\tau \hat{{\bf q}} \cdot \vec{\gamma} - \tau ( \vec{\gamma} \cdot 
\hat{{\bf p}}) (\hat{{\bf p}} \cdot \hat{{\bf q}}) \nonumber \\ 
& & \mbox{\hspace*{5ex}} \simeq 
\hat{{Q\kern-0.1em\raise0.3ex\llap{/}\kern0.15em\relax}}_\tau + \tau 
\hat{{\bf q}} \cdot \vec{\gamma} \nonumber \\ 
& & \mbox{\hspace*{5ex}} =  \gamma^0 \, . 
\end{eqnarray*} 
Then, we have 
\begin{eqnarray} 
\left( \Lambda^{a; \, \mu}_{g 1} (Q - K, Q) \right)^\ell_{j i} 
& \simeq & \frac{g^2}{\pi^2} (-)^{i + j + \ell} \left( \frac{N}{2} - 
C_F \right) T^a \, T \, \gamma^0 \nonumber \\ 
& & \times \sum_{\tau = \pm} \hat{Q}_\tau^\mu \, {\cal 
S}_\tau^{j \ell \ell i} \, , 
\label{lambda-10} 
\end{eqnarray} 
where 
\begin{eqnarray*} 
{\cal S}^{i j k \ell}_\tau & \equiv & \int d p \, p^2 \int 
\frac{d p_0}{p_0} \, \rho_t (P) \nonumber \\ 
& & \times \int d z \, \raisebox{1.1ex}{\scriptsize{$\diamond$}} 
\mbox{\hspace{-0.33ex}} \tilde{S}^{(i j)}_\tau (Q - K + P) \, 
\raisebox{1.1ex}{\scriptsize{$\diamond$}} \mbox{\hspace{-0.33ex}} 
\tilde{S}^{(k \ell)}_\tau (Q + P) \, . 
\end{eqnarray*} 
Obviously $\tau = q_0 / |q_0|$ sector yields the leading 
contribution. ${\cal S}_\tau$s are computed in Appendix A. From 
(\ref{f6}) in Appendix A, we see that, in the region $|K \cdot 
\hat{Q}_\tau|$ $=$ $O (\Gamma_q)$, ${\cal S}_\tau$s are of $O (L / 
\Gamma_q)$ $=$ $O \{ 1 / (g^2 T) \}$ where $L$ is as in (\ref{f5}). 
Then $\left( \Lambda^{a; \, \mu}_{g 1} (Q - K, Q) 
\right)^\ell_{j i}$ in (\ref{lambda-10}) is of $O (1)$, the same 
order of magnitude as the lowest-order counterpart, $(-)^{\ell - 1} 
\delta_{\ell j} \delta_{\ell i} \, T^a \, \gamma^\mu$. It is worth 
noting that, as in the self-energy case dealt with in Sec. III, the 
dominant contribution to ${\cal S}_\tau$s comes from the region 
where $|p_0|$ $<<$ $p$ (cf. Appendix A). 

Now we substitute Fig. 2 for the quark-gluon vertex on the left side 
of Fig. 1(a) to obtain Fig. 3. Figure 3 consists of four 
contributions corresponding to $(i, j)$ $=$ $(1, 1)$, $(1, 2)$, 
$(2, 1)$, and $(2, 2)$. We first note that the form of the leading 
part of the magnetic soft gluon propagator 
$\displaystyle{\raisebox{0.6ex}{\scriptsize{*}}} \! \Delta_{\rho 
\sigma}^{\mbox{\scriptsize{mag}} (i j)} (P)$ is independent of the 
thermal indexes $i$ and $j$ (cf. (\ref{til-delta-star})). Then Fig. 
3 contains 
${\cal S}_\tau$s in the combination $\sum_{i = 1}^2 (-)^i 
{\cal S}_\tau^{j i i 1}$. From (\ref{f6}) and (\ref{f7}) in Appendix 
A, we see that the cancellation occurs in the above combination, 
\[ 
\sum_{i = 1}^2 (-)^i {\cal S}_\tau^{j i i \ell} \simeq 0 \, . 
\] 
This means that, although each of the four contributions of Fig. 3 
is of the same order of magnitude as the contribution of Fig. 1, 
cancellations occur between them and the contribution of Fig. 3 
turns out to be nonleading. 

Now let us turn to analyze multi-loop contributions. We first 
inspect the ladder diagram as depicted in Fig. 4, where solid- and 
dashed-lines stand, respectively, for quark- and gluon-propagators, 
$Q$ is hard, and $P_j$s are soft. We are interested in the behavior 
at $K \cdot \hat{Q}_\tau$ $\simeq$ $0$. As will be shown below, the 
contribution from the region where $P_j$ $(1 \leq j \leq n)$ are 
soft is of the same order of magnitude as the lowest-order 
counterpart, $- (-)^{\ell} \delta_{\ell j_1} \delta_{i_1} \, T^a \, 
\gamma^\mu$. In place of (\ref{g1-1}), we have 
\begin{eqnarray} 
& & \left( \Lambda^{a ; \, \mu}_{g n} (Q - K, Q) 
\right)^\ell_{j_1 i_1} \nonumber \\ 
& & 
\mbox{\hspace*{5ex}} 
= - (-)^\ell \left[ - i g^2 \left( \frac{N}{2} - C_F \right) 
\right]^n \, T^a \nonumber \\ 
& & 
\mbox{\hspace*{8ex}} 
\times \sum_{i_2, ..., i_n = 1}^2 \sum_{j_2, ..., j_n = 1}^2 
\int_{\mbox{\scriptsize{soft}} \, P\mbox{\scriptsize{s}}} 
\prod_{k = 1}^n \left[  \frac{d^{\, 4} P_k}{(2 \pi)^4} \, (-)^{i_k + 
j_k} \, \displaystyle{\raisebox{0.6ex}{\scriptsize{*}}} \! 
\Delta_{\xi_k \zeta_k}^{\mbox{\scriptsize{mag}} (i_k j_k)} (P_k) 
\right] \nonumber \\ 
& & 
\mbox{\hspace*{8ex}} 
\times \gamma^{\zeta_1} \, 
\raisebox{1.1ex}{\scriptsize{$\diamond$}} \mbox{\hspace{-0.33ex}} 
S^{(j_1 j_2)} (Q - K + P_1) \, \gamma^{\zeta_2} \, 
\raisebox{1.1ex}{\scriptsize{$\diamond$}} \mbox{\hspace{-0.33ex}} 
S^{(j_2 j_3)} (Q - K + \sum_{j = 1}^2 P_j) \, \gamma^{\zeta_3} \, 
\cdot \cdot \cdot \nonumber \\ 
& & 
\mbox{\hspace*{8ex}} 
\times \gamma^{\zeta_n} \, 
\raisebox{1.1ex}{\scriptsize{$\diamond$}} \mbox{\hspace{-0.33ex}} 
S^{(j_n \ell)} (Q - K + \sum_{j = 1}^n P_j) \, \gamma^\mu \, 
\raisebox{1.1ex}{\scriptsize{$\diamond$}} \mbox{\hspace{-0.33ex}} 
S^{(\ell i_n)} (Q + \sum_{j = 1}^n P_j) \, \gamma^{\xi_n} \, \cdot 
\cdot \cdot \nonumber \\ 
& & 
\mbox{\hspace*{8ex}} 
\times \gamma^{\xi_3} \, 
\raisebox{1.1ex}{\scriptsize{$\diamond$}} \mbox{\hspace{-0.33ex}} 
S^{(i_3 i_2)} (Q + \sum_{j = 1}^2 P_j) \, \gamma^{\xi_2} \, 
\raisebox{1.1ex}{\scriptsize{$\diamond$}} \mbox{\hspace{-0.33ex}} 
S^{(i_2 i_1)} (Q + P_1) \,  \gamma^{\xi_1} \, . 
\label{gn-1} 
\end{eqnarray} 
We substitute (\ref{delta-star}), (\ref{til-delta-star}), and 
(\ref{kekka-1}) into (\ref{gn-1}). We 
then carry out the $P_k$-integration successively starting from 
$P_1$-integration and then $P_2$-integration and so on. From 
(\ref{gn-1}), pick out the term 
\begin{equation} 
\int d p_k \, p_k^2 \int d p_{k 0} \, \frac{\rho_t (P_k)}{p_{k 0}} 
\int d (\hat{{\bf p}}_k \cdot \hat{{\bf q}}) \, 
\raisebox{1.1ex}{\scriptsize{$\diamond$}} \mbox{\hspace{-0.33ex}} 
\tilde{S}^{(j_k j_{k + 1})}_\tau (Q - K + \sum_{j = 1}^k P_j) \, 
\raisebox{1.1ex}{\scriptsize{$\diamond$}} \mbox{\hspace{-0.33ex}} 
\tilde{S}^{(i_{k + 1} i_k)}_\tau (Q + \sum_{j = 1}^k P_j) \, . 
\label{gn-2} 
\end{equation} 
As has been noted above (and in Appendix A), the region, from which 
the leading contribution emerges, is 
\[ 
|p_{j 0}| << p_j \, , \;\;\;\;\;\; |{\bf p}_j \cdot \hat{{\bf q}}| 
<< p_j \;\;\;\;\;\;\; (1 \leq j \leq k - 1) \, . 
\] 
Then we see that $|r_{k 0} - \tau r_k|$ $<<$ $p_k$ $=$ $O (g T)$, 
where $R_k$ $\equiv$ $Q$ $+$ $\sum_{j = 1}^{k - 1} P_j$. Thus for 
(\ref{gn-2}), we can use the result obtained in Appendix A: 
\[ 
\mbox{Eq. (\ref{gn-2})} = {\cal S}_\tau^{j_k j_{k + 1} i_{k + 1} 
i_k} \, , \;\;\;\;\;\;\;\; (i_{n + 1} = j_{n + 1} = \ell) \, . 
\] 
Using all this, we obtain 
\begin{eqnarray} 
& & \left( \Lambda^{a ; \, \mu}_{g n} (Q - K, Q) \right)^\ell_{j_1 
i_1} \nonumber \\ 
& & \mbox{\hspace*{5ex}} =  - (-)^{i_1 + j_1 + \ell} \left[ - 
\frac{g^2}{\pi^2} \left( \frac{N}{2} - C_F \right) \, T \right]^n 
\nonumber \\ 
& & 
\mbox{\hspace*{18ex}} \times T^a \, \hat{Q}_\tau^\mu \, \gamma^0 \, 
{\cal S}_\tau^{(n) j_1 \ell \ell i_1} \, , 
\label{gn-5} 
\end{eqnarray} 
where 
\begin{eqnarray*} 
& & {\cal S}_\tau^{(n) i j k \ell} \equiv \sum_{i_1, \, j_1 = 1}^2 
(-)^{i_1 + j_1} {\cal S}_\tau^{(n - 1) i j_1 i_1 \ell} \, {\cal 
S}_\tau^{j_1 j k i_1} \, , \nonumber \\ 
& & \mbox{\hspace*{15ex}} \left( {\cal S}_\tau^{(0) i j k \ell} 
\equiv {\cal S}_\tau^{i j k \ell} \right) \, . 
\end{eqnarray*} 
Using the formulas in Appendix A, we can show by induction that 
\[ 
{\cal S}^{(n) i j k \ell}_\tau = \left( \frac{i \tau \pi}{2} \, L 
\right)^n \sum_{\sigma = \pm} \frac{\sigma^{n -1} \, 
a_\tau^{(\sigma) i j k \ell}}{\left( K \cdot \hat{R}_\tau + 2 i 
\sigma \tau \Gamma \right)^n} \, , 
\] 
where $a_\tau^{(\sigma)}$s are as in (\ref{f7}) in Appendix A. As in 
the case of $n = 1$, in the region $|K \cdot \hat{Q}_\tau|$ $=$ $O 
(\Gamma_q)$, $\left( \Lambda^{a ; \, \mu}_{g n} (Q - K, Q) \right)$s 
in (\ref{gn-5}) is of $O (1)$. 

Now we inspect Fig. 5, which is obtained from Fig. 1 by replacing 
the bare quark-gluon vertexes with their multi-loop \lq\lq 
corrections'', Fig. 4. Each contribution $\tilde{\Sigma}_F^{j_1 j_2 
i_2 i_1}$ is of the same order of magnitude as the contribution of 
Fig. 1. As in the case of $n = 1$ above, from the formulas in 
Appendix A, we can readily derive 
\[ 
\sum_{i = 1}^2 (-)^i \, {\cal S}_\tau^{(n) j i i \ell} \simeq 0 \, . 
\] 
Then, upon summation over $j_1$, $j_2$, $i_1$, and $i_2$, 
cancellation occurs between the contributions of Fig. 5 and the 
whole contribution, $\sum_{j_1, \, j_2, \, i_2, \, i_1 = 1}^2 
\tilde{\Sigma}_F^{j_1 j_2 i_2 i_1}$, turns out to be nonleading. 

We now turn to analyze a \lq\lq crossed-ladder'' diagram, which is 
obtained from Fig. 4 by interchanging the vertexes with thermal 
indexes $j_1, \, j_2$, ..., $j_n$. Since the leading part of a 
soft-gluon propagator, Eqs. (\ref{til-delta-star}), is 
independent of the 
thermal indexes, the analysis of the \lq\lq crossed-ladder'' diagram 
does not bring about any complexity as compared to the ladder 
diagram Fig. 4. Then, Fig. 5 with \lq\lq crossed-ladder'' diagram(s) 
leads to nonleading $\tilde{\Sigma}_F$. 
\subsection{Analysis of Figs. 6 and 7} 
There is yet another diagram, Fig. 6, that leads to $O (1)$ 
contribution to the quark-gluon vertex. In Fig. 6, $Q$ is hard and 
$P$ {\em and} $K$ are soft. 

Fig. 6 yields 
\begin{eqnarray} 
\left( \Lambda^{a ; \, \mu}_{\mbox{\scriptsize{Fig. 6}}} (Q - K, Q) 
\right)^\ell_{j i} & = & - \frac{i}{2} \, (-)^{i + j} \, g^2 \, N \, 
T^a \int_{\mbox{\scriptsize{soft}} \, P} \frac{d^{\, 4} 
P}{(2 \pi)^4} \, \gamma_\sigma \, 
\raisebox{1.1ex}{\scriptsize{$\diamond$}} \mbox{\hspace{-0.33ex}} 
S^{(j i)} (Q + P) \, \gamma_\rho \nonumber \\ 
& & \times \displaystyle{\raisebox{0.6ex}{\scriptsize{*}}} \! 
\Delta^{\rho \xi (i i')} (P) \, 
\displaystyle{\raisebox{0.6ex}{\scriptsize{*}}} \! \Delta^{\zeta 
\sigma (j' j)} (K + P) \, \left( 
\displaystyle{\raisebox{0.6ex}{\scriptsize{*}}} {\cal V}^\mu_{\xi 
\zeta} (K, P) \right)^\ell_{i' j'} \, . 
\label{3-glu} 
\end{eqnarray} 
Here 
\begin{eqnarray*} 
\left( \displaystyle{\raisebox{0.6ex}{\scriptsize{*}}} {\cal 
V}^\mu_{\xi \zeta} (K, P) \right)^\ell_{i' j'} & = & \left( 
{\cal V}^{(0) ; \, \mu}_{\xi \zeta} (K, P) \right)^\ell_{i' j'} 
\nonumber \\ 
& & + \left( \displaystyle{\raisebox{0.6ex}{\scriptsize{*}}} 
\tilde{{\cal V}}^\mu_{\xi \zeta} (K, P) \right)^\ell_{i' j'} 
\nonumber \\  
\left( {\cal V}^{(0) ; \, \mu}_{\xi \zeta} (K, P) \right)^\ell_{i' 
j'} & = & (-)^{\ell - 1} \, \delta^\ell_{i'} \, \delta^\ell_{j'} 
\left\{ \delta^\mu_\xi (K - P)_\zeta \right. \nonumber \\ 
& & \left. + g_{\xi \zeta} (K + 2 P)^\mu - \delta^\mu_\zeta (2 K + 
P)_\xi \right\} \, , 
\end{eqnarray*} 
where $\displaystyle{\raisebox{0.6ex}{\scriptsize{*}}} 
\tilde{{\cal V}}^\mu_{\xi \zeta}$ is the HTL-contribution. 

Let us estimate the order of magnitude of (\ref{3-glu}) in the 
region (\ref{ryouiki}), $||q_0| - q|$ $\leq$ $O \{ g^2 T \ln 
(g^{- 1}) \}$. We ignore possible factors of $\ln (g^{- 1})$ and 
keep only powers of $g$. From (\ref{kekka-4}), we have 
\[ 
\displaystyle{ \raisebox{0.9ex}{\scriptsize{$\diamond$}}} 
\mbox{\hspace{-0.33ex}} \rho_\tau (Q + P) \simeq \frac{1}{\pi} \, 
\frac{\gamma_q}{[q_0 - \tau q + p_0 - \tau {\bf p} \cdot \hat{{\bf 
q}}]^2 + \gamma_q^2} \, , 
\] 
where $\gamma_q \simeq \gamma_q (Q) = O (g^2 T)$. Then, when $|p_0 - 
\tau {\bf p} \cdot \hat{{\bf q}}|$ $=$ $O (g^2 T)$, $\displaystyle{ 
\raisebox{0.9ex}{\scriptsize{$\diamond$}}} \mbox{\hspace{-0.33ex}} 
\rho_\tau (Q + P)$ is of $O \{ 1 / (g^2 T) \}$. This is also the 
case for $\raisebox{1.1ex}{\scriptsize{$\diamond$}} 
\mbox{\hspace{-0.33ex}} S^{(j i)} (Q + P)$. Thus, $\int d^{\, 4} P$ 
$=$ $\int d p_0 \int d p \, p^2 \int d (\hat{{\bf p}} \cdot 
\hat{{\bf q}})$ $=$ $O (g^5 T^4)$. 
$\displaystyle{\raisebox{0.6ex}{\scriptsize{*}}} \! \Delta$s are of 
$O \{ 1 / (g^3 T^2) \}$ and 
$\displaystyle{\raisebox{0.6ex}{\scriptsize{*}}} {\cal 
V}^\mu_{\xi \zeta}$ is of $O (g T)$. Collecting all of them, we 
have 
\[ 
\left( \Lambda^{a ; \, \mu}_{\mbox{\scriptsize{Fig. 6}}} (Q - K, Q) 
\right)^\ell_{j i} = O (1) \, , 
\] 
which is of the same order of magnitude as the bare quark-gluon 
vertex. 

Let us turn to inspect Fig. 7, which is obtained from Fig. 1(a) by 
replacing the left bare quark-gluon vertex with Fig. 6. The same 
\lq\lq phenomenon'' as in the case of Fig. 3 occurs here. Each 
contribution of Fig. 7, which corresponds to a set of values of $i, 
j, i'$, and $j'$, is of the same order of magnitude as the 
contribution of Fig. 1. Noting again that all the soft-gluon 
propagator are independent of the thermal indexes and recalling the 
identity $\sum_{i, \, i', \, j' = 1}^2 \left( 
\displaystyle{\raisebox{0.6ex}{\scriptsize{*}}} {\cal V}^\mu_{\xi 
\zeta} (K, P) \right)^\i_{i i' j'}$ $=$ $0$, we see that, upon 
summation over $i$, $i'$, and $j'$ in Fig. 7, cancellation takes 
place between the contributions. Thus the contribution of Fig. 7 is 
nonleading. 
\setcounter{equation}{0}
\setcounter{section}{4}
\section{Hard-gluon mode} 
\def\theequation{\mbox{\arabic{section}.\arabic{equation}}}
The analysis goes parallel to that of Sec. III, so that we briefly 
present. The region of our interest is (\ref{region}) or more 
precisely (\ref{ryouiki}). 

In Appendix B, computation of the one-loop contribution to $Re \, 
\tilde{\Pi}_F^{\nu \mu} (Q)$ and $Re \, \tilde{\Pi}_F^g (Q)$is 
carried out in conventional hot QCD. The resultant $Re \, 
\tilde{\Pi}_F$s are 
\begin{eqnarray} 
Re \, \tilde{\Pi}_F^T (Q) \rule[-4mm]{0.2mm}{9mm} 
\raisebox{-3.4mm}{\scriptsize{\hspace{0.5mm}one-loops}} & \simeq & 
\frac{3}{2} \, m_T^2 \, , 
\label{Re-pi-t} \\ 
Re \, \tilde{\Pi}_F^L (Q) \rule[-4mm]{0.2mm}{9mm} 
\raisebox{-3.4mm}{\scriptsize{\hspace{0.5mm}one-loops}} & \simeq & 
Re \, \tilde{\Pi}_F^C (Q) \rule[-4mm]{0.2mm}{9mm} 
\raisebox{-3.4mm}{\scriptsize{\hspace{0.5mm}one-loops}} \nonumber \\ 
& \simeq & 0 \, , 
\label{Re-pi-l} \\ 
Re \, \tilde{\Pi}_F^D (Q) \rule[-4mm]{0.2mm}{9mm} 
\raisebox{-3.4mm}{\scriptsize{\hspace{0.5mm}one-loops}} & = & 0 \, , 
\label{Re-pi-d} \\ 
Re \, \tilde{\Pi}_F^g (Q) \rule[-4mm]{0.2mm}{9mm} 
\raisebox{-3.4mm}{\scriptsize{\hspace{0.5mm}one-loops}} & \simeq & 0 
\, ,  
\label{Re-pi-g} 
\end{eqnarray} 
where $m_T$ is as in (\ref{gluon-mass}). The above results are gauge 
independent. Eqs. (\ref{Re-pi-l}) and (\ref{Re-pi-g}) mean that $|Re 
\, \Pi_F^A (Q)|$ $<<$ $(g T)^2$ $(A$ $=$ $L$, $C$, $g)$. 

As in the case of hard-quark self-energy part, to leading order, the 
theory defined by (\ref{2.1a}) - (\ref{qf-FP}) and (\ref{int-1}) 
yields the 
same results, (\ref{Re-pi-t}) - (\ref{Re-pi-g}) (cf. Appendix B). 

The self-consistency conditions $\Pi^{\mu \nu}_F (Q)$ $=$ 
$\tilde{\Pi}^{\mu \nu}_F (Q)$ and $\Pi_F^g (Q)$ $=$ $\tilde{\Pi}_F^g 
(Q)$ with $\Pi_F^{\mu \nu} (Q)$ in (\ref{qf-gluon}) and $\Pi_F^g 
(Q)$ in (\ref{qf-FP}) yield 
\begin{eqnarray} 
Re \, \Pi_F^T (Q) & \simeq & \frac{3}{2} \, m_T^2 \, , 
\label{maru-a} \\ 
Re \, \Pi_F^L (Q) & \simeq & Re \, \Pi_F^C (Q) \simeq Re \, \Pi_F^g 
(Q) \nonumber \\ 
& \simeq & 0 \, . 
\label{maru-b} \\ 
Re \, \Pi_F^D (Q) & = & 0 \, , 
\label{maru-c} 
\end{eqnarray} 
which meet the requirements of BRS invariance, (\ref{piD0}), 
(\ref{WT-4}), and (\ref{WT-5}). 

Let us turn to analyze $Im \, \tilde{\Pi}_F (Q)$. It has been proved 
\cite{kobes} that, in conventional hot QCD, the pole positions of 
the transverse and longitudinal propagators (cf. (\ref{g-p-t}) and 
(\ref{g-p-l})) are independent of the choice of gauge. The diagram 
that yields the leading contribution to $Im \, \tilde{\Pi}_F^{\mu 
\nu} (Q)$ is depicted in Fig. 8(a), where $K$ is soft. 

Let us first compute the contribution from Fig. 8 to $Im \, 
\tilde{\Pi}_F^T (Q)$ and $Im \, \tilde{\Pi}_F^L (Q)$ in our theory. 
The contribution from Fig. 8(b) is 
\[ 
Im \, \tilde{\Pi}_F^{A \, (8 b)} (Q) = - Im \, \Pi_F^A (Q) \;\;\;\;\;
\;\;\; (A = T, L) \, . 
\] 
For calculating Fig. 8(a), as in \cite{pis5}, for calculational 
ease, we use Coulomb gauge, in which only the transverse part of the 
hard-gluon propagator $\raisebox{1.1ex}{\scriptsize{$\diamond$}} 
\mbox{\hspace{-0.33ex}} \Delta^{\rho \sigma} (Q - K)$ contributes. 

The leading contribution to $Im \, \tilde{\Pi}_F^T (Q)$ comes from 
Fig. 8(a) with magnetic part of the soft-gluon propagator. The same 
remark above after (\ref{til-delta-star}) applies here. 
Straightforward 
calculation yields 
\begin{eqnarray*} 
Im \, \tilde{\Pi}_F^{T \, (8 a)} (Q) & \simeq & 2 g^2 \, N \, q \, T 
\int_{\mbox{\scriptsize{soft}} \, K} \frac{d^{\, 4} K}{(2 \pi)^3} \, 
\frac{\rho_t (K)}{k_0} \{ 1 - (\hat{{\bf q}} \cdot \hat{{\bf 
k}})^2\} \\ 
& & \times \frac{g_T (q_0, q)}{\left[ q_0 - \tau \left( q + {\cal 
M}_T^2 / q \right) - k_0 + \tau {\bf k} \cdot \hat{{\bf q}} 
\right]^2 + [g_T (q_0, q)]^2} \, , 
\end{eqnarray*} 
where $\tau = q_0 / |q_0|$, ${\cal M}_T$ is as in 
(\ref{small-k0}), and 
\[ 
g_T (q_0, q) = \frac{1}{2 q} \, Im \, \Pi_F^T (q_0, q) \, . 
\] 
Proceeding as in Sec. III, we obtain 
\begin{equation} 
Im \, \tilde{\Pi}_F^{T \, (8 a)} \simeq \frac{g^2}{2 \pi} \, N \, q 
\, T \, \frac{g_T (q_0, q)}{|g_T (q_0, q)|} \, \ln \left( 
\frac{m_T}{\Gamma_T (Q)} \right) \, , 
\label{8a-2} 
\end{equation} 
where 
\[ 
\Gamma_T (Q) \equiv max \left[ |q_0 - \tau \left( q + {\cal M}_T^2 / 
q \right)|, \; |g_T (q_0, q)| \right] \, . 
\] 
Eq. (\ref{8a-2}) is valid at logarithmic accuracy. In conventional 
hot QCD, we have (\ref{8a-2}) with $g_T / |g_T|$ $=$ $-1$ and 
$\Gamma_T$ $=$ $|q_0 - \tau q|$, which is gauge independent. Then 
(\ref{8a-2}) is also gauge independent. 

The self-consistency condition yields, in place of (\ref{quark-2}) 
in Sec. III, 
\begin{eqnarray*} 
Im \, \Pi_F^{T \, (8 a)} (q_0, q) & = & 2 q \, g_T (q_0, q) \equiv - 
2 q \, \gamma_T \, , \\ 
\gamma_T & = & \frac{g^2}{4 \pi} \, N \, T \, \ln (g^{- 1}) \\ 
& & \times \left[ 1 - \frac{\ln \{ \ln (g^{- 1}) \}}{\ln (g^{- 1})} 
+ {\cal F} \right] + O (g^2 T) \, . 
\end{eqnarray*} 

The contribution to $Im \, \tilde{\Pi}_F^{L \, (8 a)}$ from Fig. 
8(a) reads 
\begin{eqnarray*} 
& & Im \, \tilde{\Pi}_F^{L \, (8 a)} \nonumber \\ 
& & 
\mbox{\hspace*{5ex}} 
= - 2 g^2 \, N \, 
\frac{T}{q} 
\int_{\mbox{\scriptsize{soft}} \, K} \frac{d^{\, 4} K}{(2 \pi)^3} 
\frac{g_T (q_0, q)}{\left[ q_0 - \tau (q + {\cal M}_T^2 / q) - k_0 
+ \tau {\bf k} \cdot \hat{{\bf q}} \right]^2 + [g_T (q_0, q)]^2} \\ 
& & 
\mbox{\hspace*{8ex}} 
\times \left[ \frac{\rho_t (K)}{k_0} \left\{ 1 + [ \hat{{\bf k}} 
\cdot (\widehat{{\bf q} - {\bf k}}) ]^2 \right\} \right. \\ 
& & 
\mbox{\hspace*{8ex}} 
\times \left\{ k_0^2 - (\hat{{\bf q}} \cdot {\bf k})^2 + \frac{2 
(Q - K)^2 - 2 K^2 - Q^2}{4} - \frac{\{ (Q - K)^2 - K^2 \}^2}{4 Q^2} 
\right\} \\ 
& & 
\mbox{\hspace*{8ex}} 
+ \frac{\rho_\ell (K)}{k_0} \left\{ \left( q_0 - \frac{k_0}{2} 
\right)^2 \left[ 1 - \left\{ \hat{{\bf q}} \cdot (\widehat{{\bf q} - 
{\bf k}}) \right\}^2 \right] \right. \\ 
& & \left. \left. 
\mbox{\hspace*{8ex}} 
- \left( 1 - \frac{k_0^2}{4 Q^2} \right) \left[ 
k^2 - \left\{ {\bf k} \cdot (\widehat{{\bf q} - {\bf k}}) \right\}^2 
\right] \right\} \right] \, , 
\end{eqnarray*} 
where $\rho_\ell (K)$ is the spectral function of the 
electric part of the soft-gluon propagator. Simple dimensional 
analysis yields 
\begin{eqnarray} 
Im \, \tilde{\Pi}_F^{L \, (8 a)} & = & Im \, \Pi_F^L (Q) \nonumber 
\\ 
& = & O (g^4 T^2) + Q^2 \times O (g^2) + \frac{1}{Q^2} \times O 
(g^6 T^4) \, , 
\label{8a-11} 
\end{eqnarray} 
where factors of $\ln (g^{- 1})$ are ignored. The first equality is 
due to the self-consistency condition. 

The contribution from Fig. 8(a) to $Im \, \tilde{\Pi}_F^L$, where 
both $K$ {\em and} $R - K$ are hard, may be analyzed similarly. The 
order of magnitude of the resultant contribution is again given by 
(\ref{8a-11}). 

Eq. (\ref{8a-11}) shows that, at $|Q^2|$ $=$ $O (g^3 T^2)$, $|Im \, 
\Pi_F^L (Q)|$ $=$ $O (g^3 T^2)$, which is of the same order of 
magnitude as $|Q^2|$. In the region, 
\begin{equation} 
|Q^2| >> g^3 T^2 \, , 
\label{wakareme} 
\end{equation} 
$|Im \, \Pi_F^L| << |Q^2|$, which means that the resummation of $Im 
\, \Pi_F^L$ is not necessary. Thus, in the region (\ref{wakareme}), 
we have 
\[ 
Im \, \Pi_F^L (Q) \simeq 0 \, . 
\] 
This together with (\ref{maru-b}) yields 
\begin{eqnarray} 
\Pi_F^L (Q) & \simeq & 0 \, , 
\label{kaku-1} \\ 
\raisebox{1.1ex}{\scriptsize{$\diamond$}} \mbox{\hspace{-0.33ex}} 
\Delta_F^L (Q) & \simeq & \frac{1}{Q^2 + i 0^+} \, . 
\label{kaku} 
\end{eqnarray} 
Since this form is valid in the region (\ref{wakareme}), $+ i 0^+$ 
in the denominator is not necessary. Nevertheless we have kept it 
for the reason to be discussed below. As mentioned above, pole 
position of $\raisebox{1.1ex}{\scriptsize{$\diamond$}} 
\mbox{\hspace{-0.33ex}} \Delta_F^L (Q)$ is independent of the choice 
of gauge. Then the form (\ref{kaku}) is gauge independent. 

Eqs. (\ref{WT-4}) and (\ref{WT-5}) with (\ref{kaku-1}) and 
(\ref{maru-b}) yield 
\begin{equation} 
\Pi_F^C (Q) \simeq \Pi_F^g (Q) \simeq 0 \, , 
\label{kaku-3} 
\end{equation} 
which together with (\ref{g-p-c}) and (\ref{FP-prop}) leads to 
\begin{eqnarray} 
\raisebox{1.1ex}{\scriptsize{$\diamond$}} \mbox{\hspace{-0.33ex}} 
\Delta_F^C (Q) \simeq 0 \, , 
\label{kaku-4} \\ 
\raisebox{1.1ex}{\scriptsize{$\diamond$}} \mbox{\hspace{-0.33ex}} 
\Delta_F^g (Q) \simeq \frac{1}{Q^2 + i 0^+} \, . 
\label{kaku-5} 
\end{eqnarray} 
The forms (\ref{kaku-3}) - (\ref{kaku-5}) are valid in the region 
(\ref{wakareme}). 

Summarizing the result obtained above, we have for the diagonalized 
gluon- and FP-ghost-propagators, 
\begin{eqnarray} 
\raisebox{1.1ex}{\scriptsize{$\diamond$}} \mbox{\hspace{-0.33ex}} 
\Delta_F^T (Q) & = & \frac{1}{Q^2 - 3 m_T^2 / 2 - i Im \, \Pi_F^T 
(Q)} \nonumber \\ 
& \simeq &  \frac{\epsilon (q_0)}{2 q} \, \frac{1}{q_0 - \epsilon 
(q_0) (q + {\cal M}_T^2 / q) + i \epsilon (q_0) \, \gamma_T} \, , 
\label{8a-13-1} \\ 
\raisebox{1.1ex}{\scriptsize{$\diamond$}} \mbox{\hspace{-0.33ex}} 
\Delta_F^L (Q) & \simeq & \raisebox{1.1ex}{\scriptsize{$\diamond$}} 
\mbox{\hspace{-0.33ex}} \Delta_F^g (Q) \nonumber \\ 
& \simeq & \frac{1}{Q^2 + i 0^+} \, , 
\label{8a-14-2} \\ 
\raisebox{1.1ex}{\scriptsize{$\diamond$}} \mbox{\hspace{-0.33ex}} 
\Delta_F^C (Q) & \simeq  & 0 \, , 
\label{8a-13-3} \\ 
\raisebox{1.1ex}{\scriptsize{$\diamond$}} \mbox{\hspace{-0.33ex}} 
\Delta_F^D (Q) & = & \eta \, \frac{1}{Q^2 + i 0^+} \, . 
\label{8a-13-4} 
\end{eqnarray} 
The $2 \times 2$ gluon propagator is related to (\ref{8a-13-1}) - 
(\ref{8a-13-4}) through the relations (\ref{mat-1}) - (\ref{mat-3}). 

Here the similar observation as that at the end of Sec. III applies. 
The form of $Im \, \Delta_F^T (Q)$ is valid in the region 
(\ref{ryouiki}). The forms of $Re \, \Delta_F^T (Q)$, $\Delta_F^L 
(Q)$, $\Delta_F^g (Q)$, and $\Delta_F^C (Q)$ are valid in the region 
$O (g^3 T)$ $<$ $|q_0 - \epsilon (q_0) (q + {\cal M}_T^2 / q)|$ 
$\leq$ $O \{ g^2 T \ln (g^{- 1}) \}$. For evaluating $Re \, 
\Delta_F^T (Q)$, $\Delta_F^L (Q)$, $\Delta_F^g (Q)$, and $\Delta_F^C 
(Q)$ in the region $|q_0 - \epsilon (q_0) (q + {\cal M}_T^2 / q)|$ 
$\leq$ $O (g^3 T)$, concrete evaluation of one-loop diagram as well 
as of two-loop diagram is necessary. 

Suppose that we calculate some thermal amplitude. We encounter the 
integral, 
\[ 
\sum_{A = T, \, L, \, C, \, D, \, g} \int d^{\, 4} Q \, \Delta_F^{A 
\, (r s)} (Q) \, {\cal H}^A (Q) \;\;\;\;\;\;\;\;\; (r, s = 1, 2) \, 
. 
\] 
If this integral is insensitive to the region $|Q^2|$ $\leq$ $O (g^3 
T^2)$, we can use (\ref{8a-14-2}) for $\Delta_F^L (Q)$ {\em and} 
$\Delta_F^g (Q)$ and (\ref{8a-13-3}) for $\Delta_F^C (Q)$. This is 
because the phase-space volume of the region $|Q^2|$ $\leq$ $O (g^3 
T^2)$ is $O \{g / \ln (g^{- 1}) \}$ smaller than the volume of the 
region $O (g^3 T^2)$ $\leq$ $|Q^2|$ $\leq$ $O \{ g^2 T^2 \ln (g^{- 
1}) \}$. In the opposite case, we cannot use (\ref{8a-14-2}) and 
(\ref{8a-13-3}) and, as stated above, the analysis including the 
next-to-leading order is necessary. 
\section*{Acknowledgment} 
This work was supported in 
part by the Grant-in-Aide for Scientific Research ((A)(1) (No. 
08304024)) of the Ministry of Education, Science and Culture of 
Japan. 
\setcounter{equation}{0}
\setcounter{section}{1}
\section*{Appendix A} 
\def\theequation{\mbox{\Alph{section}.\arabic{equation}}} 
In this Appendix, we compute ${\cal S}_\tau^{i j k \ell}$: 
\begin{eqnarray} 
{\cal S}^{i j k \ell}_\tau & \equiv & \int d p \, p^2 \int 
\frac{d p_0}{p_0} \, \rho_t (P) \nonumber \\ 
& & \times \int d z \, \raisebox{1.1ex}{\scriptsize{$\diamond$}} 
\mbox{\hspace{-0.33ex}} \tilde{S}^{(i j)}_\tau (R - K + P) \, 
\raisebox{1.1ex}{\scriptsize{$\diamond$}} \mbox{\hspace{-0.33ex}} 
\tilde{S}^{(k \ell)}_\tau (R + P) \, , 
\label{f1} 
\end{eqnarray} 
where, $z$ $\equiv$ $\tau \hat{{\bf p}} \cdot \hat{{\bf r}}$. In 
(\ref{f1}), $R$ is hard and $P$ {\em and} $K$ are soft. We are 
interested in the form of ${\cal S}_\tau$s in the region where $|r_0 
- \tau r|$ $=$ $O (\Gamma_q)$ $=$ $O \{ g^2 T \ln (g^{- 1}) \}$ 
($r_0$ $=$ $\tau |r_0|$) and $|K \cdot \hat{R}_\tau|$ $<<$ $g T$. 

${\cal S}^{1 1 1 1}_\tau$ reads 
\begin{eqnarray*} 
{\cal S}^{1 1 1 1}_\tau & = & \int d p \, p^2 \int \frac{d p_0}{p_0} 
\, \rho_t (P) \int_{- 1}^1 d z \, 
\raisebox{1.1ex}{\scriptsize{$\diamond$}} \mbox{\hspace{-0.33ex}} 
\tilde{S}^{(1 1)}_\tau (R - K + P) \, 
\raisebox{1.1ex}{\scriptsize{$\diamond$}} \mbox{\hspace{-0.33ex}} 
\tilde{S}^{(1 1)}_\tau (R + P) \\ 
& \simeq & \frac{1}{4} \int d p \, p^2 \int \frac{d p_0}{p_0} \, 
\rho_t (P) \int_{- 1}^1 d z \, \left[ \frac{1 - n_F}{r_0 - \tau r + 
p_0 - p z - K \cdot \hat{R}_\tau + i \tau \Gamma_q} \right. \\ 
& & \left. + \frac{n_F}{r_0 - \tau r + p_0 - p z - K \cdot 
\hat{R}_\tau - i \tau \Gamma_q} \right] \nonumber \\ 
& & \times \left[ \frac{1 - n_F}{r_0 - \tau r + p_0 - p z + i \tau 
\Gamma_q} + \frac{n_F}{r_0 - \tau r + p_0 - p z - i \tau \Gamma_q} 
\right] \\ 
& \simeq & \frac{1}{4} \int d p \, p \int \frac{d p_0}{p_0} \, 
\rho_t (P) \nonumber \\ 
& & \times \sum_{\sigma = \pm} \left[ \frac{n_F (1 - n_F)}{K \cdot 
\hat{R}_\tau + 2 i \sigma \tau \Gamma_q} \left\{ \ln \left( \frac{- 
p + p_0 + r_0 - \tau r + i \sigma \tau \Gamma_q}{p + p_0 + r_0 - 
\tau r + i \sigma \tau \Gamma_q} \right) \right. \right. \\  
& & \left. + \ln \left( \frac{p + p_0 + r_0 - \tau r - K \cdot 
\hat{R}_\tau - i \sigma \tau \Gamma_q}{- p + p_0 + r_0 - \tau r - K 
\cdot \hat{R}_\tau - i \sigma \tau \Gamma_q} \right) \right\} \\ 
& & + \frac{ \{ \theta (- \sigma) - n_F \}^2}{K \cdot \hat{R}_\tau} 
\left\{ \ln \left( \frac{- p + p_0 + r_0 - \tau r - i \sigma \tau 
\Gamma_q}{p + p_0 + r_0 - \tau r - i \sigma \tau \Gamma_q} \right) 
\right. \\ 
& & \left. \left. + \ln \left( \frac{p + p_0 + r_0 - \tau r - K 
\cdot \hat{R}_\tau - i \sigma \tau \Gamma_q}{- p + p_0 + r_0 - \tau 
r - K \cdot \hat{R}_\tau - i \sigma \tau \Gamma_q} \right) \right\} 
\right] \, , 
\end{eqnarray*} 
where $n_F$ $\equiv$ $n_F (q)$. The quantity in the second curly 
brackets leads to nonleading contribution. As in the case of 
self-energy part in Sec. III, at logarithmic accuracy, the region 
$|p_0|$ $<<$ $p$ yields the dominant contribution. Then we drop the 
factors $p_0$ in the arguments of logarithms. Using (\ref{small-k0}) 
and integrating over $- \beta p$ $<$ $p_0$ $<$ $+ \beta p$ with 
$\beta << 1$, we obtain 
\begin{eqnarray*} 
{\cal S}^{1 1 1 1}_\tau & = & \frac{1}{2 \pi} \int \frac{d p}{p} \, 
\arctan \left( \frac{\pi \beta {\cal M}_T^2}{p^2} \right) \nonumber 
\\ 
& & \times \sum_{\sigma = \pm} \left[ \frac{n_F (1 - n_F)}{K \cdot 
\hat{R}_\tau +  2 i \sigma \tau \Gamma_q} \left\{ \ln \left( \frac{- 
p + p_0 + r_0 - \tau r + i \sigma \tau \Gamma_q}{p + p_0 + r_0 - 
\tau r + i \sigma \tau \Gamma_q} \right) \right. \right. \\  
& & \left. \left. + \ln \left( \frac{p + p_0 + r_0 - \tau r - K 
\cdot \hat{R}_\tau - i \sigma \tau \Gamma_q}{- p + p_0 + r_0 - \tau 
r - K \cdot \hat{R}_\tau - i \sigma \tau \Gamma_q} \right) \right\} 
\right] \, .  
\end{eqnarray*} 
Let us inspect the first logarithmic function with $p_0 = 0$, 
$\ln (...)$. For $p$ $=$ $O (g T)$, $\ln (...)$ $\simeq$ $i \sigma 
\tau \pi$, $\ln (...)$ $= 0$ at $p = 0$, and the transition region 
is $p$ $=$ $O (\Gamma_q)$. Similar observation may be made for the 
second logarithmic function.   

Then, we obtain, at logarithmic accuracy, 
\begin{equation} 
{\cal S}^{1 1 1 1}_\tau \simeq \frac{i \tau \pi}{2} \, L 
\sum_{\sigma = \pm} \frac{\sigma \, n_F (1 - n_F)}{K \cdot 
\hat{R}_\tau + 2 i \sigma \tau \Gamma_q} \, ,  
\label{f4} 
\end{equation} 
where 
\begin{equation} 
L \equiv \frac{1}{2} \left[ \ln \left( \frac{m_T}{\Gamma_q} \right) 
+ \ln \left( \frac{m_T}{\mbox{max} \left( \Gamma_q, \, |K \cdot 
\hat{R}_\tau| \right)} \right) \right] \, . 
\label{f5} 
\end{equation} 
It is to be noted that, at logarithmic accuracy, the restriction of 
the $z$ region to $|z|$ $<<$ $1$ or $|\hat{{\bf p}} \cdot \hat{{\bf 
r}}|$ $<<$ $p$ yields the same result (\ref{f4}). 

In a similar manner, we can compute all ${\cal S}^{i j k 
\ell}_\tau$. Writing 
\begin{equation} 
{\cal S}^{i j k \ell}_\tau \simeq \frac{i \tau \pi}{2} \, L 
\sum_{\sigma = \pm} \frac{a^{(\sigma) i j k \ell}_\tau}{K \cdot 
\hat{R}_\tau + 2 i \sigma \tau \Gamma_q} \, , 
\label{f6} 
\end{equation} 
we have  
\begin{eqnarray} 
a^{(\pm) 1 1 1 1}_\tau & = & a_\tau^{(\pm) 2 2 2 2} = a^{(\pm) 1 2 
2 1}_\tau = a^{(\pm) 2 1 1 2}_\tau \nonumber \\ 
& = & \pm n_F [1 - n_F] \, , \nonumber \\ 
a^{(\pm) 1 1 1 2}_\tau & = & a^{(\pm) 1 1 2 1}_{- \tau} = - a^{(\mp) 
2 2 2 1}_{- \tau} = - a^{(\mp) 2 2 1 2}_\tau \nonumber \\ 
& = & - a^{(\mp) 1 2 1 1}_\tau = - a^{(\mp) 2 1 1 1}_{- \tau} = 
a^{(\pm) 2 1 2 2}_{- \tau} = a^{(\pm) 1 2 2 2}_\tau \nonumber \\ 
& = & \mp [ \theta (- \tau) - n_F] [ \theta (\mp) - n_F ] \, , 
\nonumber \\ 
a^{(\pm) 1 1 2 2}_\tau & = & - a^{(\mp) 2 2 1 1}_\tau = \mp [\theta 
(\mp) - n_F]^2 \nonumber \\ 
a^{(\pm) 1 2 1 2}_\tau & = & a^{(\pm) 2 1 2 1}_{- \tau} \nonumber \\ 
& = & \mp [ \theta (- \tau) - n_F]^2 \, . 
\label{f7} 
\end{eqnarray} 
\setcounter{equation}{0}
\setcounter{section}{2}
\section*{Appendix B Computation of the real part of 
$\tilde{\Pi}^{\nu \mu}_F (Q)$} 
\def\theequation{\mbox{\Alph{section}.\arabic{equation}}} In this 
Appendix, we compute the real part of the one-loop thermal gluon 
self-energy part, $Re \, \tilde{\Pi}_F^{(q) \, \nu \mu} (Q)$, in 
conventional hot QCD. The region of our interest is 
\begin{equation} 
|Q^\mu| = O (T) \, , \;\;\;\;\;\; |Q^2| \leq O \{ g^2 T^2 \ln 
(g^{- 1}) \} \, . 
\label{b1} 
\end{equation} 
\subsection*{1. Contribution of a quark loop} 
The contribution of a quark loop reads 
\begin{equation} 
Re \, \tilde{\Pi}_F^{(q) \, \nu \mu} (Q) = - \frac{g^2}{2} \, N_f \, 
Re \int \frac{d^{\, 4} K}{(2 \pi)^4} \, tr \left[ \gamma^\mu \, 
S^{(1 1)} (K - Q) \, \gamma^\nu \, S^{(1 1)} (K) \right] \, . 
\label{tsuika1} 
\end{equation} 
After straightforward manipulations, we have for $Re \, 
\tilde{\Pi}_F^{(q) \, T} (Q)$, 
\begin{eqnarray} 
Re \,\tilde{\Pi}_F^{(q) \, T} (Q) & = & \frac{g^2}{12} \, N_f T^2 
\nonumber \\ 
& & - \frac{g^2}{4 \pi^2} \, N_f \, \frac{Q^2}{q^3} \int_0^\infty d 
k \, k^2 n_F (k) \sum_{\tau = \pm} \ln \left( \frac{|Q^2 + 2 q k - 2 
\tau q_0 k|}{|Q^2 - 2 q k - 2 \tau q_0 k|} \right) \nonumber \\ 
& & + \frac{g^2}{8 \pi^2} \, N_f \, \frac{Q^2}{q^2} \int_0^\infty d 
k \, n_F (k) \nonumber \\ 
& & \times \left[ 4 k - \sum_{\tau = \pm} \frac{2 q^2 + Q^2 - 4 
\tau q_0 k}{2 q} \ln \left( \frac{|Q^2 + 2 q k - 2 \tau q_0 k|}{|Q^2 
- 2 q k - 2 \tau q_0 k|} \right) \right] \, . \nonumber \\ 
\label{b2} 
\end{eqnarray} 
It can easily be shown that, in the region (\ref{b1}), no $O (g^2 
T^2)$ contribution emerges from the second and third terms on the 
R.H.S. Then we have 
\begin{equation} 
Re \, \tilde{\Pi}_F^{(q) \, T} (Q) \simeq \frac{g^2}{12} \, N_f \, 
T^2 \, . 
\label{b3} 
\end{equation} 
We next have 
\begin{eqnarray} 
Re \, \left( \tilde{\Pi}_F^{(q) \, L} (Q) + 2 \tilde{\Pi}_F^{(q) \, 
T} (Q) \right) & = & Re \, \Pi_F^{(q) \, \nu \mu} (Q) \left( g_{\mu 
\nu} - \frac{Q_\mu Q_\nu}{Q^2} \right) \nonumber \\ 
& = & \frac{g^2}{6} \, N_f \, T^2 \nonumber \\ 
& & - \frac{g^2}{4 \pi^2} \, N_f \, \frac{Q^2}{q} \int_0^\infty d k 
\, n_F (k) \nonumber \\ 
& & \times \sum_{\tau = \pm} \ln \left( \frac{|Q^2 + 2 q k - 2 \tau 
q_0 k|}{|Q^2 - 2 q k - 2 \tau q_0 k|} \right) \nonumber \\ 
& \simeq & \frac{g^2}{6} \, N_f \, T^2 \, . 
\label{b4} 
\end{eqnarray} 

>From (\ref{b3}) and (\ref{b4}), we obtain 
\begin{equation} 
Re \, \tilde{\Pi}_F^{(q) \, L} (Q) \simeq 0 \, . 
\label{b5} 
\end{equation} 

$Re \, \tilde{\Pi}_F^{(q) \, \nu \mu} (Q)$ in (\ref{tsuika1}) 
satisfies $Q_\nu \, \tilde{\Pi}_F^{(q) \, \nu \mu} (Q)$ $=$ $0$, so 
that 
\[ 
Re \, \tilde{\Pi}_F^{(q) \, C} (Q) = Re \, \tilde{\Pi}_F^{(q) \, D} 
(Q) = 0 \, . 
\] 
It is to be noted that the integrals in (\ref{b2}) and (\ref{b4}) 
are insensitive to the region $K \simeq 0$ or $Q - K \simeq 0$. Then 
the replacement $S^{(1 1)} (K)$ $\to$ 
$\displaystyle{\raisebox{0.6ex}{\scriptsize{*}}} \! S^{(11)} (K)$ 
[$S^{(1 1)} (Q - K)$ $\to$ 
$\displaystyle{\raisebox{0.6ex}{\scriptsize{*}}} \! S^{(1 1)} (Q - 
K)$] in the soft-$K$ [soft-(Q - K)] region in (\ref{tsuika1}) does 
not bring about the contribution of $O (g^2 T^2)$. 

Also to be noted is that the leading contributions, (\ref{b3}) and 
(\ref{b5}), have come from the hard-$K$ {\em and} hard-$(Q - K)$ 
region. From the above derivation, we can easily verify that the the 
formula (\ref{tsuika1}) with 
$\raisebox{1.1ex}{\scriptsize{$\diamond$}} \mbox{\hspace{-0.33ex}} 
S^{(1 1)}$s for $S^{(1 1)}$s leads to the same leading-order results 
(\ref{b3}) and (\ref{b5}). 
\subsection*{2. Contribution of gluon loops and a FP-ghost loop} 
The $(1, 1)$-component of the bare thermal gluon propagator is 
written as 
\begin{eqnarray} 
\Delta_{\mu \nu}^{(1 1)} (Q) & = & - \left[ g_{\mu \nu} - (\eta - 1) 
\, Q_\mu \, Q_\nu \frac{\partial}{\partial \lambda^2} \right] 
\nonumber \\ 
& & \times \Delta^{(1 1)} (Q ; \lambda^2) \rule[-3mm]{.14mm}{8.5mm} 
\raisebox{-2.85mm}{\scriptsize{$\; \lambda = 0$}} \, , 
\mbox{\hspace{7.5ex}} (j, \, \ell = 1, 2) \, , \nonumber 
\\ 
\Delta^{(11)} (Q ; \lambda^2) & = & \frac{1}{Q^2 - \lambda^2 + i 
0^+} \nonumber \\ 
& & - 2 \pi i n_B (|q_0|) \, \delta (Q^2 - \lambda^2) \, , \nonumber 
\end{eqnarray} 
where $n_B (x) \equiv 1 / (e^{x / T} - 1)$. Accordingly the 
contribution to $\tilde{\Pi}_F^{\nu \mu}$ of gluon loops plus the 
contribution of FP-ghost loop consists of three terms, 
\begin{eqnarray} 
\tilde{\Pi}^{\nu \mu}_F (Q) & = & \tilde{\Pi}^{\nu \mu \, (0)}_F (Q) 
+ (\eta - 1)\, \tilde{\Pi}^{\nu \mu \, (1)}_F (Q) \nonumber \\ 
& & + (\eta - 1)^2 \, \tilde{\Pi}^{\nu \mu \, (2)}_F (Q) \, , 
\nonumber 
\end{eqnarray} 
and the one-loop contribution to the FP-ghost self-energy part 
$\tilde{\Pi}_F^g (Q)$ consists of two terms 
\begin{equation} 
\tilde{\Pi}_F^g (Q) = \tilde{\Pi}_F^{g \, (0)} (Q) + (\eta - 1) \, 
\tilde{\Pi}_F^{g \, (1)} (Q) \, . 
\label{eta-FP} 
\end{equation} 
We summarize the result of the straightforward calculation. 
$\tilde{\Pi}^{(j) \nu \mu}_F (Q)$ $(j = 0, 1, 2)$ satisfies the 
relation, 
\[ 
Q_\nu \, \tilde{\Pi}^{\nu \mu \, (0)}_F (Q) = Q_\nu \, 
\tilde{\Pi}^{\nu \mu \, (2)}_F (Q) = Q_\nu Q_\mu \, 
\tilde{\Pi}_F^{\nu \mu \, (1)} (Q) = 0 \, , 
\] 
so that, with obvious notations, 
\begin{eqnarray*} 
Re \, \tilde{\Pi}_F^{C \, (0)} & = & Re \, \tilde{\Pi}_F^{D \, (0)} 
= Re \, \tilde{\Pi}_F^{D \, (1)} = Re \, \tilde{\Pi}_F^{C \, (2)} \\ 
& & = Re \, \tilde{\Pi}_F^{D \, (2)} = 0 \, . 
\end{eqnarray*} 
Nonvanishing $Re \, \tilde{\Pi}_F$s are 
\begin{eqnarray} 
& & Re \, \tilde{\Pi}_F^{T \, (0)} (Q) \nonumber \\ 
& & \mbox{\hspace*{4ex}} = \frac{g^2}{6} \, N \, T^2 - \frac{g^2}{4 
\pi^2} \, N \, \frac{Q^2}{q^3} \int_0^\infty  dk \, k^2 \, n_B (k) 
\sum_{\tau = \pm} \ln \left( \frac{|Q^2 + 2 q k - 2 \tau q_0 
k|}{|Q^2 - 2 q k - 2 \tau q_0 k|} \right) \nonumber \\ 
& & \mbox{\hspace*{6.5ex}} + \frac{g^2}{8 \pi^2} \, N \, 
\frac{Q^2}{q^2} \int_0^\infty dk \, n_B (k) \nonumber \\ 
& & \mbox{\hspace*{6.5ex}} \times \left[ 4 k - \sum_{\tau 
= \pm} \frac{4 q^2 + Q^2 - 4 \tau q_0 k}{2 q} \, \ln \left( 
\frac{|Q^2 + 2 q k - 2 \tau q_0 k|}{|Q^2 - 2 q k - 2 \tau q_0 k|} 
\right) \right] \, , 
\label{b21} \\ 
& & Re \, \left( \tilde{\Pi}_F^{L \, (0)} (Q) + 2 \tilde{\Pi}_F^{T 
\, (0)} (Q) \right) \nonumber \\ 
& & \mbox{\hspace*{4ex}} = \frac{g^2}{3} \, N \, T^2 - \frac{5}{8 
\pi^2} \, g^2 \, N \, \frac{Q^2}{q} \int_0^\infty d k \, n_B (k) 
\sum_{\tau = \pm} \ln \left( \frac{|Q^2 + 2 q k - 2 \tau q_0 
k|}{|Q^2 - 2 q k - 2 \tau q_0 k|} \right) \, , \nonumber \\ 
\label{b22} \\ 
& & Re \, \tilde{\Pi}_F^{T \, (1)} (Q) \nonumber \\ 
& & \mbox{\hspace*{4ex}} = \frac{g^2}{16 \pi^2} \, N \, 
\frac{Q^2}{q} \, \frac{\partial}{\partial \mu^2} \int_0^\infty d k 
\, \frac{k}{\sqrt{k^2 + \mu^2}} \, n_B (\sqrt{k^2 + \mu^2}) 
\nonumber \\ 
& & \mbox{\hspace*{6.5ex}} \times \sum_{\tau = \pm} \left[ \left\{ 
Q^2 \left( 1 + \frac{k^2}{q^2} \right) + \frac{(q^2 + \mu^2)^2 - 4 
\tau q_0 \sqrt{k^2 + \mu^2} (Q^2 + \mu^2) + 4 q_0^2 \mu^2}{4 q^2} 
\right\} \right. \nonumber \\ 
& & \left. \mbox{\hspace*{6.5ex}} \times L_\tau (\mu^2 , \lambda^2 = 
0) - 4 q k - \frac{k}{q} \, (Q^2 + \mu^2) \right] 
\rule[-3mm]{.14mm}{8.5mm} \raisebox{-2.85mm}{\scriptsize{$\; \mu^2 
= 0$}} \nonumber \\ 
& & \mbox{\hspace*{6.5ex}} + \frac{g^2}{16 \pi^2} \, N \, 
\frac{Q^2}{q} \, \frac{\partial}{\partial \lambda^2} \int_0^\infty d 
k \, n_B (k) \nonumber \\ 
& & \mbox{\hspace*{6.5ex}} \times \sum_{\tau = \pm} \left[ \left\{ 
Q^2 \left( 1 + \frac{k^2}{q^2} \right) + \frac{(Q^2 - \lambda^2) 
(Q^2 - \lambda^2 - 4 \tau q_0 k)}{4 q^2} \right\} \right. \nonumber 
\\ 
& & \mbox{\hspace*{6.5ex}} \left. \times L_\tau (\mu^2 = 0 , 
\lambda^2) - \frac{k}{q} \, (Q^2 - \lambda^2) \right] 
\rule[-3mm]{.14mm}{8.5mm} \raisebox{-2.85mm}{\scriptsize{$\; 
\lambda^2 = 0$}} \, , 
\label{b23} \\ 
& & Re \, \left( \tilde{\Pi}_F^{L \, (1)} (Q) + 2 \tilde{\Pi}_F^{T 
\, (1)} (Q) \right) \nonumber \\ 
& & \mbox{\hspace*{4ex}} = \frac{5 g^2}{32 \pi^2} \, N \, 
\frac{Q^2}{q} \, \frac{\partial}{\partial \mu^2} \int_0^\infty d k 
\, \frac{k}{\sqrt{k^2 + \mu^2}} \, n_B (\sqrt{k^2 + \mu^2}) 
\nonumber \\ 
& & \mbox{\hspace*{6.5ex}} \times \sum_{\tau = \pm} \left[ \left( 
Q^2 + \frac{2}{5} \, \mu^2 \right) L_\tau (\mu^2, \lambda^2 = 0) - 4 
q k \right] \rule[-3mm]{.14mm}{8.5mm} 
\raisebox{-2.85mm}{\scriptsize{$\; \mu^2 = 0$}} \nonumber \\ 
& & \mbox{\hspace*{6.5ex}} + \frac{5 g^2}{32 \pi^2} \, N \, 
\frac{Q^2}{q} \, \frac{\partial}{\partial \lambda^2} \int_0^\infty d 
k \, n_B (k) \left( Q^2 + \frac{2}{5} \, \lambda^2 \right) 
\sum_{\tau = \pm} L_\tau (\mu^2 = 0 , \lambda^2) 
\rule[-3mm]{.14mm}{8.5mm} \raisebox{-2.85mm}{\scriptsize{$\; 
\lambda^2 = 0$}} \, , \nonumber \\ 
\label{b24} \\ 
& & Re \, \tilde{\Pi}_F^{C \, (1)} (Q) \nonumber \\ 
& & \mbox{\hspace*{4ex}} = - \frac{\sqrt {2} \, g^2}{32 \pi^2} \, N 
\, \frac{Q^2}{q^2} \, \frac{\partial}{\partial \mu^2} \int_0^\infty 
d k \, \frac{k}{\sqrt{k^2 + \mu^2}} \, n_B (\sqrt{k^2 + \mu^2}) 
\nonumber \\ 
& & \mbox{\hspace*{6.5ex}} \times \left[ 4 k q q_0 + Q^2 \sum_{\tau 
= \pm} \left( \tau \sqrt{k^2 + \mu^2} - \frac{q_0}{2} \right) \, 
L_\tau (\mu^2, \lambda^2 = 0) \right] \rule[-3mm]{.14mm}{8.5mm} 
\raisebox{-2.85mm}{\scriptsize{$\; \mu^2 = 0$}} \nonumber \\ 
& & \mbox{\hspace*{6.5ex}} + \frac{\sqrt {2} \, g^2}{32 \pi^2} \, N 
\, \frac{Q^4}{q^2} \, \frac{\partial}{\partial \lambda^2} 
\int_0^\infty d k \, n_B (k) \sum_{\tau = \pm} \left( \tau k - 
\frac{q_0}{2} \right) \, L_\tau (\mu^2 = 0 , \lambda^2) 
\rule[-3mm]{.14mm}{8.5mm} \raisebox{-2.85mm}{\scriptsize{$\; 
\lambda^2 = 0$}} \, , \nonumber \\ 
& & \mbox{\hspace*{6.5ex}} + \frac{\sqrt {2} \, g^2}{32 \pi^2} \, N 
\, Q^2 \, \frac{q_0}{q^2} \int_0^\infty d k \, n_B (k) \, L_\tau 
(\mu^2 = 0, \lambda^2 = 0) \, , 
\label{b24-1} \\ 
& & Re \, \tilde{\Pi}_F^{T \, (2)} (Q) \nonumber \\ 
& & \mbox{\hspace*{4ex}} = \frac{g^2}{32 \pi^2} \, N \, 
\frac{Q^4}{q^3} \, \frac{\partial}{\partial \mu^2} \, 
\frac{\partial}{\partial \lambda^2} \int_0^\infty d k \, 
\frac{k}{\sqrt{k^2 + \mu^2}} \, n_B (\sqrt{k^2 + \mu^2}) \nonumber 
\\ 
& & \mbox{\hspace*{6.5ex}} \times \sum_{\tau = \pm} \left[ (Q^2 + 
\mu^2 - \lambda^2) q k \right. \nonumber \\ 
& & \mbox{\hspace*{6.5ex}} - \frac{4 k^2 Q^2 + (Q^2 + \mu^2 - 
\lambda^2)^2 - 4 \tau q_0 \sqrt{k^2 + \mu^2} (Q^2 + \mu^2 - 
\lambda^2) + 4 q_0^2 \mu^2}{4} \nonumber \\ 
& & \left. \mbox{\hspace*{6.5ex}} \times L_\tau (\mu^2, \lambda^2) 
\right] \rule[-3mm]{.14mm}{8.5mm} \raisebox{-2.85mm}{\scriptsize{$\; 
\mu^2 = \lambda^2 = 0$}} \, , 
\label{b25} \\ 
& & Re \, \left( \tilde{\Pi}_F^{L \, (2)} (Q) + 2 \tilde{\Pi}_F^{T 
\, (2)} (Q) \right) \nonumber \\ 
& & \mbox{\hspace*{4ex}} = \frac{g^2}{64 \pi^2} \, N \, 
\frac{Q^4}{q} \, \frac{\partial}{\partial \mu^2} \, 
\frac{\partial}{\partial \lambda^2} \int_0^\infty d k \, 
\frac{k}{\sqrt{k^2 + \mu^2}} \, n_B (\sqrt{k^2 + \mu^2}) \nonumber 
\\ 
& & \mbox{\hspace*{6.5ex}} \times \sum_{\tau = \pm} \left[ - 4 q k 
\, \frac{Q^2 + \mu^2 - \lambda^2}{Q^2} + \{ Q^2 - 2 (\mu^2 + 
\lambda^2) \} L_\tau (\mu^2 , \lambda^2) \right] 
\rule[-3mm]{.14mm}{8.5mm} \raisebox{-2.85mm}{\scriptsize{$\; \mu^2 = 
\lambda^2 = 0$}} \, . \nonumber \\ 
\label{b26} 
\end{eqnarray} 
Here 
\[ 
L_\tau (\mu^2, \lambda^2) \equiv \ln \left( \frac{|Q^2 + 2 q k - 2 
\tau q_0 \sqrt{k^2 + \mu^2} + \mu^2 - \lambda^2|}{|Q^2 - 2 q k - 2 
\tau q_0 \sqrt{k^2 + \mu^2} + \mu^2 - \lambda^2|} \right) \, . 
\] 

$\tilde{\Pi}_F^{g \, (j)} (Q)$ $(j = 0, 1)$ in (\ref{eta-FP}) reads 
\begin{eqnarray} 
Re \, \tilde{\Pi}_F^{g \, (0)} (Q) & = & - \frac{g^2}{16 \pi^2} \, N 
\, \frac{Q^2}{q} \int_0^\infty dk \, n_B (k) \sum_{\tau = \pm} \ln 
\left( \frac{|Q^2 + 2 q k - 2 \tau q_0 k|}{|Q^2 - 2 q k - 2 \tau q_0 
k|} \right) \nonumber \\ 
\label{b30} \\ 
Re \, \tilde{\Pi}_F^{g \, (1)} (Q) & = & \frac{g^2}{64 \pi^2} \, N 
\, \frac{Q^2}{q} \, \frac{\partial}{\partial \mu^2} \int_0^\infty d 
k \, \frac{k}{\sqrt{k^2 + \mu^2}} \, n_B (\sqrt{k^2 + \mu^2}) 
\nonumber \\ 
& & \times \sum_{\tau = \pm} \left[ Q^2 L_\tau (\mu^2, \lambda^2 = 
0) - 4 q k \right] \rule[-3mm]{.14mm}{8.5mm} 
\raisebox{-2.85mm}{\scriptsize{$\; \mu^2 = 0$}} \nonumber \\ 
& & + \frac{g^2}{64 \pi^2} \, N \, \frac{Q^4}{q} \, 
\frac{\partial}{\partial \lambda^2} \int_0^\infty d k \, n_B (k) 
\sum_{\tau = \pm} L_\tau (\mu^2 = 0 , \lambda^2) 
\rule[-3mm]{.14mm}{8.5mm} \raisebox{-2.85mm}{\scriptsize{$\; 
\lambda^2 = 0$}} \, . \nonumber \\ 
\label{b31} 
\end{eqnarray} 

As in the previous subsection 1, we can easily see that no $O 
(g^2 T^2)$ contribution arises from the integrals in (\ref{b21}), 
(\ref{b22}), and (\ref{b30}). Computation of (\ref{b23}) - 
(\ref{b26}) and (\ref{b31}) goes as follows. Take derivative with 
respect to $\mu^2$ and/or $\lambda^2$ and set $\mu^2$ $=$ 
$\lambda^2$ $=$ $0$. Divide the integration region into $0$ $\leq$ 
$k$ $\leq$ $k^*$ and $k^*$ $\leq$ $k$, where $g T$ $<$ $k^*$ $<$ 
$T$. It can readily be shown that the contributions from the latter 
region are nonleading when compared to (\ref{b3}) and (\ref{b4}). 
The contributions from the former region may be calculated 
explicitly by using $n_B (k)$ $\simeq$ $T / k$ and $n_F (k)$ 
$\simeq$ $1 / 2$ and are shown to be also nonleading. 

After all this, we have 
\begin{eqnarray} 
Re \, \tilde{\Pi}_F^T (Q) & \simeq & \frac{g^2}{6} \, N \, T^2 \, , 
\label{b210} \\ 
Re \, \tilde{\Pi}_F^L (Q) & \simeq & \tilde{\Pi}_F^C (Q) \simeq 
\tilde{\Pi}_F^g (Q) \simeq 0 \, , 
\label{b211} \\ 
Re \, \tilde{\Pi}_F^D (Q) & = & 0 \, , 
\label{b212} 
\end{eqnarray} 
which are gauge independent. Similar observation as that at the end 
of subsection 1 applies here. 

Using (\ref{b3}), (\ref{b5}), and (\ref{b210}) - (\ref{b212}), we 
finally obtain 
\begin{eqnarray*} 
Re \, \tilde{\Pi}_F^T (Q) & \simeq & \frac{1}{6} \, \left( N + 
\frac{N_f}{2} \right) \, (g T)^2 \;\; \left( = \frac{3}{2} \, m_T^2 
\right) \, , \\ 
Re \, \tilde{\Pi}_F^L (Q) & \simeq & Re \, \tilde{\Pi}_F^C (Q) 
\simeq Re \, \tilde{\Pi}_F^g (Q) \simeq 0 \, . \\ 
Re \, \tilde{\Pi}_F^D (Q) & = & 0 \, . 
\end{eqnarray*} 

\newpage 
\begin{center} 
{\Large {\bf Figure captions} } \vspace*{1.5em} 
\end{center} 
\begin{description} 
\item[Fig. 1.] Diagrams for the self-energy part of the quark mode. 
\item[Fig. 2.] An one-loop diagram for the quark-gluon vertex. 
\lq\lq $\ell$'', \lq\lq $i$'', and \lq\lq $j$'' are thermal indexes. 
The blob on the gluon line indicates the effective soft-gluon 
propagator. 
\item[Fig. 3.] A two-loop diagram for the self-energy part of the 
quark mode. 
\item[Fig. 4.] An $n$-loop diagram for the quark-gluon vertex. 
\lq\lq $\ell$'', \lq\lq $i_1$'' $-$ \lq\lq $i_n$'' and \lq\lq 
$j_1$'' $-$ \lq\lq $j_n$'' are thermal indexes. 
\item[Fig. 5.] A multi-loop diagram for the self-energy part of the 
quark mode. 
\item[Fig. 6.] 
An one-loop diagram for the quark-gluon vertex. The 
blob on the vertex indicates the effective soft tri-gluon vertex. 
\item[Fig. 7.] A two-loop diagram for the self-energy part of the 
quark mode. 
\item[Fig. 8.] Diagrams for the self-energy part of the gluon mode. 
\end{description} 
\end{document}